\documentclass[aip,jcp,reprint]{revtex4-1}
\usepackage{amssymb,amsmath}
\usepackage{graphicx}

\begin{document}

\title{Interacting hard rods on a lattice:
  Distribution of microstates and density functionals}

\author{Benaoumeur Bakhti} 
\affiliation{Fachbereich Physik, Universit\"at Osnabr\"uck,
  Barbarastra{\ss}e 7, 49076 Osnabr\"uck, Germany}

\author{Gerhard M\"uller} 
\affiliation{Department of Physics, University of Rhode Island,
  Kingston RI 02881, USA}

\author{Philipp Maass}
\email{philipp.maass@uni-osnabrueck.de}
\affiliation{Fachbereich Physik, Universit\"at Osnabr\"uck,
Barbarastra{\ss}e 7, 49076 Osnabr\"uck, Germany}

\date{3 May 2013}

\begin{abstract}
  We derive exact density functionals for systems of hard rods with
  first-neighbor interactions of arbitrary shape but limited range on
  a one-dimensional lattice. The size of all rods is the same integer
  unit of the lattice constant. The derivation, constructed from
  conditional probabilities in a Markov chain approach, yields the
  exact joint probability distribution for the positions of the rods
  as a functional of their density profile. For contact interaction
  (``sticky core model'') between rods we give a lattice fundamental
  measure form of the density functional and present explicit results
  for contact correlators, entropy, free energy, and chemical
  potential. Our treatment includes inhomogeneous couplings and
  external potentials.
\end{abstract}



\maketitle

\section{Introduction}
\label{sec:introduction}

Lattice density functional theory is drawing increasing attention on
account of its wide range of applicability to phenomena of strong
interest in current research.\cite{Nieswand/etal:1993a,
  Reinel/Dieterich:1996, Reinhard/etal:2000} Its applications include
ordering phenomena in metallic alloys, submonolayer adsorbate
systems,\cite{Gouyet/etal:2003} colloid-polymers mixtures,
\cite{Cuesta/etal:2005} fluids in porous media,
\cite{Kierlik/etal:2001} and DNA denaturation.\cite{Azbel:1979} In
certain cases, exact density functionals in zero and one dimension can
be written in a so-called fundamental measure form, which allows one
to construct approximate functionals in higher dimensions that reduce
to the exact ones upon dimensional crossover. \cite{Rosenfeld:1989,
  Lafuente/Cuesta:2002b, Schmidt/etal:2003, Lafuente/Cuesta:2003a,
  Lafuente/Cuesta:2003b, Lafuente/Cuesta:2004} Moreover, the theory
can be extended to time-dependent
phenomena. \cite{Heinrichs/etal:2004, Kessler/etal:2002,
  Dierl/etal:2011, Dierl/etal:2012}

Particles with shapes that interact solely via hard-core repulsion on
a lattice or in a continuum do produce interesting effects including
phase transitions, e.g.\ for hard hexagons.\cite{Baxter:1980} However,
the inclusion of attractive or repulsive forces on contact or at some
distance is important for more realistic modelling.\cite{Percus:1982,
  Brannock/Percus:1996, Choudhury/Ghosh:1997, Acedo/Santos:2001,
  Gazzillo/etal:2004, Miller/Frenkel:2004a, Miller/Frenkel:2004b,
  Rickayzen/Heyes:2007, Buzzaccaro/etal:2007, Santos/etal:2008,
  Hansen-goos/Wettlaufer:2011, Hansen-goos/etal:2012} Models with
square-well interaction potentials including potentials of the
zero-range, sticky-core type have proven to exhibit realistic physical
features for many systems: colloidal suspensions,\cite{Jamnik:1998,
  Pontoni/etal:2003, Lajovic/etal:2009} crystallization of polymers,
\cite{Dasmahapatra/etal:2009, Hoy/OHern:2010} micelles,
\cite{Amokrane/Regnaut:1997} protein solutions,
\cite{Lomakin/etal:1996, Braun:2002} DNA coated colloids
\cite{Xu/etal:2011, Dreyfus/etal:2009} ionic fluids\cite{Stell:1995},
and microemulsions.\cite{Seto/etal:2000, Robertus/etal:1989} It is
well established that many effects of generic short-range interactions
can be reproduced by contact forces of a strength that yields matching
second virial coefficients.\cite{Gazzillo/etal:2006}

\begin{figure}[b!]
\includegraphics[width=0.4\textwidth]{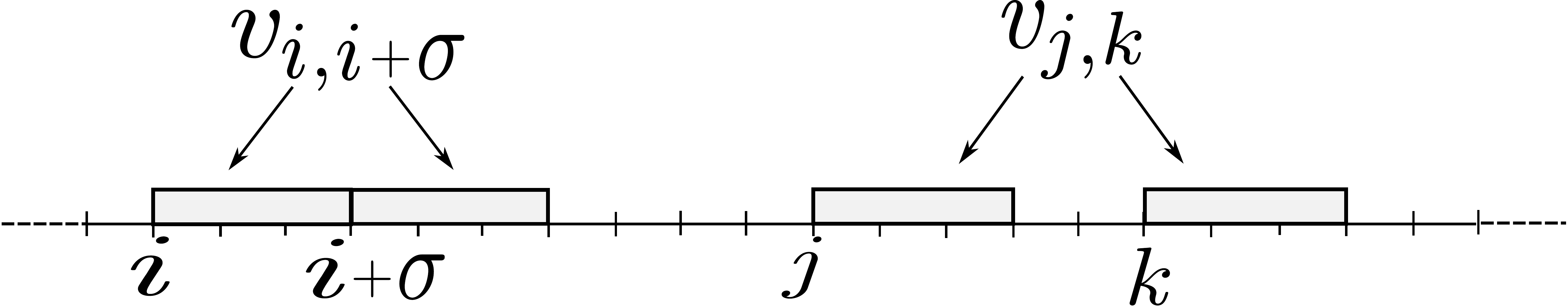}
\caption{\label{fig:model} Hard rods of size $\sigma=3$ interacting
  with a potential of range $\xi<2\sigma$. The rods
  on the left are in contact. The interaction between the two rods is
  nonzero for $k-j\le\xi$.}
 \label{fig:fig1}
\end{figure}

In this paper we develop ideas found in the work of Buschle \textit{et
  al.} \cite{Buschle/etal:2000a, Buschle/etal:2000b} to derive an
exact free-energy functional for hard rods with first-neighbor
coupling of arbitrary shape and limited range.  Onto a one-dimensional
lattice of $L$ sites we place non-overlapping rods of one size
$\sigma$ (in unit of the lattice spacing) as illustrated in
Fig.~\ref{fig:fig1}.  To each lattice site $i$ we assign occupation
number $n_i=1$ if it is the location of the left end of a rod and
$n_i=0$ otherwise.  The hard-core exclusion condition implies that
$n_in_{i+j}=0$ for $j=0,\ldots,\sigma-1$.  Hard walls at $i=1,L$ imply
that $n_i=0$ for $i<1$ and $i>L-\sigma$.

The two terms in the Hamiltonian,
\begin{equation}
\label{eq:hamiltonian}
  \mathcal{H}(\mathbf{n})=
  \sum\limits_{i<j} v_{i,j}n_{i}n_{j}+\sum\limits_{i}u_in_{i}\,,
\end{equation}
represent an interaction $v_{i,j}$ between successive rods and an
external potential $u_i$.  We write $\mathbf{n}=\{n_1,\ldots,n_L\}$
for microstates (for easy notation, we include the $n_i=0$ for
$i=L-\sigma+1,\ldots,L$).  The range $\xi$ of the interaction is
assumed to be shorter than two rod lengths.  Hence we have
$v_{i,j}=\infty$ for $|j-i|<\sigma$ and $v_{i,j}=0$ for $|j-i|>\xi$,
where $\sigma\leq\xi<2\sigma$.

We present explicit results for rods subject to contact forces
$(\xi=\sigma)$, in which case our model is a lattice version of the
sticky-core continuum model analyzed by
Baxter.\cite{Baxter:1968}. For
the case $\xi=\sigma=1$ our model reduces to the familiar Ising
lattice gas.  The chemical potential, which controls the average
number of rods in a grandcanonical ensemble, is conveniently absorbed
as a constant in the external potential.

In the following we derive exact joint probability distributions for
the positions of the rods on the lattice as functionals of their
density profile, from which we infer exact expressions for the
two-point functions and density functionals. Contact is made with
lattice fundamental measure theory.  We then present exact and
explicit results for the thermodynamics pertaining to models with
attractive and repulsive contact potentials. 

In connection with the Markov property underlying the Markov chain
approach, it is interesting to note that it corresponds to Kirkwood's
``superposition principle'' in the integral equation method for
evaluating distribution functions in continuum fluids.\cite{Abe:1959}
For one-dimensional continuum fluids with nearest-neighbor
interactions, analytical expressions for many-particle densities were
derived by using Laplace transform techniques and it was shown that
the Markov property (or ``superposition principle'') then becomes
exact.\cite{Salsburg/etal:1953} Our approach for lattice systems here
starts with the Markov property and by this we are able to derive
microstate distributions as functionals of the density profile.

\section{Distribution of microstates}
\label{sec:microstates}

For given interaction $V(\mathbf{n})=\sum\limits_{i<j} v_{i,j}n_in_j$,
the external potential
$U_\mathbf{p}(\mathbf{n})=\sum_iu_i[\mathbf{p}]n_i$, which yields the
set $\mathbf{p}=\{p_1,\ldots,p_L\}$ of mean occupation numbers
$p_i=\langle n_i\rangle$ (``density profile'') in equilibrium, is a
(unique) functional of $\mathbf{p}$.\cite{Mermin:1965} Our goal in
this section is to derive the corresponding joint probability
$\chi_\mathbf{p}(\mathbf{n})\propto\exp\{-V(\mathbf{n})-\tilde
U_\mathbf{p}(\mathbf{n})\}$ as functional of $\mathbf{p}$, where
$\tilde U_\mathbf{p}(\mathbf{n})=\sum_i\tilde u_i[\mathbf{p}]n_i$,
$\tilde u_i[\mathbf{p}]=u_i[\mathbf{p}]-\mu$. All energies here and in
the following are given in units of $k_{\rm B}T$.

Our derivation, inspired by the method developed in
Ref.~\onlinecite{Buschle/etal:2000a}, proceeds in three steps. In the
first step, we use a Markov property to express $\chi(\mathbf{n})$ in
terms of marginal probabilities $\phi(n_i,\ldots,n_{i-\xi})$ for sites
within range of the interactions. In the second step the latter are
expressed in terms of the $p_i$ and the correlators
$C_{i,j}\equiv\langle n_in_j\rangle$ by making use of the exclusion
constraint associated with the hardcore repulsion.  In the last step,
the $C_{i,j}=C_{i,j}[\mathbf{p}]$ are given as functional of
$\mathbf{p}$ by comparing $\chi_\mathbf{p}(\mathbf{n})$ with the
Boltzmann expression for a few simple configurations $\mathbf{n}$.
The derivation of $\chi_\mathbf{p}(\mathbf{n})$ constitutes a rare
case, where the ``Mermin potential''
$U_\mathbf{p}(\mathbf{n})=\Omega[\mathbf{p}]-\langle
V(\mathbf{n})\rangle-\ln\chi_\mathbf{p}(\mathbf{n})$ can be stated
explicitly.

\subsection{Reduction to joint probabilities of finite range}
\label{subsec:reduction}

We begin by writing $\chi(\mathbf{n})$ as a product of conditional
probabilities,
\begin{equation}
\label{eq:chi-1}
\mathit \chi(\mathbf{n}) =
\prod_{s=1}^{L} \psi(n_s|n_{s-1},\ldots,n_1)\,.
\end{equation}
Here $\psi(n_s|n_{s-1},\ldots,n_1)$ is the probability of finding the
occupation number $n_s$ at site $s$ under the condition that all
occupation numbers left to site $s$ are given. Because of the finite
interaction range and because we are dealing with a one-dimensional
system, one can prove the following Markov property,
\begin{align}
\label{eq:markov}
  \psi(n_s|n_{s-1},\ldots,n_1)&=\psi(n_s|n_{s-1},\ldots,n_{s-\xi})\nonumber\\
&=\frac{\phi(n_s,\ldots,n_{s-\xi})}{\phi(n_{s-1},\ldots,n_{s-\xi})}\,,
\end{align}
where $\phi(n_s,\ldots,n_{s-\xi})$ denotes the joint probability
of finding the set $\{n_s,\ldots,n_{s-\xi}\}$ of occupation numbers.

\subsection{Joint probabilities of finite range}
\label{subsec:jointprob}

It is useful to subdivide the range $s,\ldots,s-\xi$ of sites into two
compartments of labels, $i\in\{s,\ldots,s-\sigma+1\}$ and
$j\in\{s-\sigma\ldots,s-\xi\}$.  The hardcore exclusion dictates that
each compartment contains at most one label of a nonzero occupation
number.  If both compartments are occupied the two labels must satisfy
$i-j>\sigma-1$.  

Let us introduce the abbreviated notation, $\phi_s(0,0)$,
$\phi_s(0,1_j)$, $\phi_s(1_i,0)$, and $\phi_s(1_i,1_j)$, for the
remaining options of compartmental occupancy. The first and second
argument refer to the first range and second range , respectively. A
zero in one argument means that all occupation numbers in the
corresponding range are zero and $1_k$ in an argument means that
$n_k=1$ in the corresponding range. For example,
$\phi_s(1_i,0)=\phi(n_s\!=\!0,\ldots,n_{i+1}\!=\!0,n_i\!=\!1,
n_{i-1}\!=\!0,\ldots,n_{s-\xi}\!=\!0)$. 

We thus arrive at the following representation for the marginal
probabilities:
\begin{align}
\label{eq:phi-decomp}
\phi(n_s,\ldots,n_{s-\xi})=&
\phi_s(0,0)^{\left(1-\sum_{(i)_1}n_i\right)\left(1-\sum_{(j)_2}n_j\right)}\nonumber\\
&{}\times\prod_{(i)_1}\phi_s(1_i,0)^{n_i(1-\sum_{(j)_2}n_j)}\nonumber\\
&{}\times\prod_{(j)_2}\phi_s(0,1_j)^{n_j(1-\sum_{(i)_1}n_i)}\nonumber\\
&{}\times\prod_{(i,j)}\phi_s(1_i,1_j)^{n_in_j}\,,
\end{align}
where $(i)_1$ and $(j)_2$ refer to indices running over the first and
second range, and the specification $(i,j)$ means that the two indices
run over the set $i=s,\ldots,s-\xi+\sigma$, $j=i-\sigma,\ldots,s-\xi$.

Equation~(\ref{eq:phi-decomp}) amounts to expressing the joint
probability in terms of the $\phi_s(.,.)$, where case selections from
all possible configurations are encoded in the exponents.  For
example, for the configuration $(1_i,0)$, the associated exponent
$n_i(1-\sum_{(j)_2}n_j)$ is one for $n_i=1$ and all $n_j=0$ in the
second range, while otherwise it is zero and thus giving an
(irrelevant) factor one in Eq.~(\ref{eq:phi-decomp}). Note that,
different from a case selection by products in the exponents, e.g.\
by $n_i\prod_{(j)_2}{(1-n_j)}$ for the configuration $(1_i,0)$, we
have written sums. This is allowed because at most one occupation
number can be one in the two ranges and
$\prod_{(j)_2}{(1-n_j)}=1-\sum_{(j)_2}n_j$.

The function $\phi(n_{s-1},\ldots,n_{s-\xi})$ in the denominator of
Eq.~(\ref{eq:markov}) has a representation identical in structure but
now the range of indices indicated by $(i)_1$ refers to
$s-1,\ldots,s-\sigma+1$ and the range of indices indicated by $(i,j)$
refers to $i=s-1,\ldots,s-\sigma-1$, $j=i-\sigma,\ldots,s-\xi$.  The
range indicated by $(j)_2$ remains unchanged. The corresponding
shortened notation is indicated by a tilde, i.e. $\tilde\phi_s(0,0)$,
$\tilde\phi_s(1_i,0)$ etc. For example,
$\tilde\phi_s(1_i,0)=\phi(n_{s-1}\!=\!0,\ldots,n_{i+1}\!=\!0,n_i\!=\!1,
n_{i-1}\!=\!0,\ldots,n_{s-\xi}\!=\!0)$.

We continue by relating the functions $\phi_s$ to the occupancies
$p_i=\langle n_i\rangle$ and correlators $C_{i,j}=\langle
n_in_j\rangle$.  For this we only need basic properties and the
normalization condition:
\begin{subequations}
\label{eq:pc}
\begin{align}
C_{i,j}&=\langle n_in_j\rangle=\sum_{(k,l)}
n_in_j\phi_s(n_k,n_l)=\phi_s(1_i,1_j)\,
\label{eq:pc-a}\\
p_i&=\langle n_i\rangle=\sum_{\{\mathbf{n}\}_s}
n_i\phi_s(n_k,n_j)\nonumber\\
&=\phi_s(1_i,0)+\sum_{(j)_2}\phi_s(1_i,1_j)\,,
\label{eq:pc-b}\\
1&=\sum_{\{\mathbf{n}\}_s} \phi_s(n_i,n_j)=
\phi_s(0,0)+\sum_{(i)_1}\phi_s(1_i,0)\nonumber\\
&\hspace{6em}{}+\sum_{(j)_2}\phi_s(0,1_j)
+\sum_{(i,j)}\phi_s(1_i,1_j)\,,
\label{eq:pc-c}
\end{align}
\end{subequations}
where $\{\mathbf{n}\}_s=\{n_s,\ldots,n_{s-\xi}\}$.
From these relations we infer
\begin{subequations}
\label{eq:phi12}
\begin{align}
\phi_s(0,0)&=1-\sum_{k=s-\xi}^s
p_k+\sum_{i=s-\xi+\sigma}^{s}\sum_{j=s-\xi}^{i-\sigma} C_{i,j}
\label{eq:phi12-a}\\
\phi_s(1_i,0)&=p_i-\sum_{j=s-\xi}^{i-\sigma}C_{i,j}
\label{eq:phi12-b}\\
\phi_s(0,1_j)&=p_j-\sum_{i=j+\sigma}^sC_{i,j}
\label{eq:phi12-c}\\
\phi_s(1_i,1_j)&=C_{i,j}
\label{eq:phi12-d}
\end{align}
\end{subequations}
The corresponding expressions for the functionals $\tilde{\phi}_s$
have the $s$ as upper limit of sums replaced by $s-1$.

Hence we have reduced the joint probability $\chi(\mathbf{n})$ to a
functional of densities $p_i$ and correlators $C_{i,j}$.

\subsection{Correlators as functionals of densities}
\label{subsec:correlators}

What remains to be accomplished is to reduce the correlators to
functionals of the densities. To this end we compare
$\chi(\mathbf{n})$ calculated from above with the Boltzmann
probability for configurations with zero, one and two rods,
\begin{subequations}
\label{eq:chi-2}
\begin{align}
\chi(0_{1},\ldots,0_{L})&=\frac{1}{\mathcal{Z}}\,,
\label{eq:chi-2-a}\\
\chi(0_{1},\ldots,1_{i},\ldots,0_{L})&=\frac{e^{-\tilde{u}_i}}{\mathcal{Z}}\,,
\label{eq:chi-2-b}\\
\chi(0_{1},\ldots,1_{j},\ldots,1_{i},\ldots,0_{L})&=
\frac{1}{\mathcal{Z}}e^{-(\tilde{u}_i+\tilde{u}_j+v_{i,j})}\,,
\label{eq:chi-2-c}
\end{align}
\end{subequations}
where $\tilde{u}_{i}=u_i-\mu$, and $i-j\ge\sigma$ in
Eq.~(\ref{eq:chi-2-c}).  These probabilites thus satisfy the relation
\begin{align}
\label{eq:chi-rel}
\frac{\chi(0_{1},\ldots,0_{L})\chi(0_{1},\ldots,1_{i},\ldots,1_{j},\ldots,0_{L})}%
{\chi(0_{1},\ldots,1_{i},\ldots,0_{L})
\chi(0_{1},\ldots,1_{j},\ldots,0_{L})}=e^{-v_{i,j}}
\end{align}

The task ahead is cumbersome but manageable: express all four joint
probabilities of (\ref{eq:chi-rel}) in terms of the marginal
probabilities $\phi_s$ and $\tilde{\phi}_s$ by using
Eqs.~(\ref{eq:chi-1}), (\ref{eq:markov}), and (\ref{eq:phi-decomp}).
Then substitute relations (\ref{eq:phi12}) in order to extract the
desired functional dependence of the correlators $C_{i,j}$ on the
densities $p_i$. 

For example, for a configuration with two rods at site $i$ and $j$
with $\sigma\le (i-j)\le\xi$ we have
\begin{align}
&\chi(0_{1},\ldots,1_{i},\ldots,1_{j},\ldots,0_{L})=\nonumber\\
&\hspace{2em}
\left[\prod_{s=1}^{j-1}\frac{\phi_s(0,0)}{\tilde\phi_s(0,0)}\right]
\frac{\phi_j(1_j,0)}{\tilde\phi_j(0,0)}
\left[\prod_{s=j+1}^{i-1}\frac{\phi_s(1_j,0)}{\tilde\phi_s(1_j,0)}\right]
\frac{\phi_i(1_i,1_j)}{\tilde\phi_j(0_i,1_j)}\nonumber\\
&\hspace{1em}{}\times
\left[\prod_{s=i+1}^{j+\xi}\frac{\phi_s(1_i,1_j)}{\tilde\phi_s(1_i,1_j)}\right]
\left[\prod_{s=j+\xi+1}^{i+\sigma-1}\frac{\phi_s(1_i,0)}{\tilde\phi_s(1_i,0)}
\right]\nonumber\\
&\hspace{1em}{}\times
\left[\prod_{s=i+\sigma}^{i+\xi}\frac{\phi_s(0,1_i)}{\tilde\phi_s(0,1_i)}
\right]
\left[\prod_{s=i+l+1}^{L}\frac{\phi_s(0,0)}{\tilde\phi_s(0,0)}\right]\,.
\end{align}
The different terms in this equation are a consequence of the Markov
chain in Eq.~(\ref{eq:markov}) with progressing $s$ index: The first
four factors arise from successively capturing the rods located at
sites $j$ and $i$. The next four terms are associated with the
following specific $s$-values: If $s=j+\xi+1$, the rod at site $j$
falls out of the interaction range, and if $s=i+\sigma$, the rod at
site $i$ is no longer in the first range (see above) with respect to
site $s$. If $s=i+\xi+1$, the rod at site $i$ eventually falls out of
the interaction range.

Decomposing the other joint probabilities in Eq.~(\ref{eq:chi-2}) in
an analogous way and inserting the $\phi(.,.)$, $\tilde\phi(.,.)$ from
Eq.~(\ref{eq:phi12}), yields, after elementary algebra,
\begin{equation}
\label{eq:c-p-rel}
C_{i,j}=\frac{\phi_i(1_i,0)\phi_i(0,1_j)}{\phi_i(0,0)}
\left[\prod_{s=i+1}^{j+\xi}\frac{\phi_s(0,1_j)}{\tilde\phi_s(0,1_j)}
\frac{\tilde\phi_s(0,0)}{\phi_s(0,0)}\right]e^{-v_{i,j}}
\end{equation}
for $\sigma\le|i-j|\le\xi$. With
the $C_{i,j}$ determined from Eqs.~(\ref{eq:c-p-rel}) as functional of
$\mathbf{p}$, the distribution of microstates becomes
also a functional of $\mathbf{p}$ using Eqs.~(\ref{eq:phi12}),
(\ref{eq:phi-decomp}), (\ref{eq:markov}), and (\ref{eq:chi-1}), i.e.\
$\chi=\chi_\mathbf{p}(\mathbf{n})$.

For general interactions $v_{i,j}$, an analytic solution
Eqs.~(\ref{eq:c-p-rel}) appears out of reach and we must resort to a
numerical evaluation.  In the remainder of this section we focus on a
system with contact interactions $v_c\equiv v_{i,i+\sigma}$. In this
case Eqs.~(\ref{eq:c-p-rel}) simplify into
\begin{equation}\label{eq:c-p-contact}
 C_{i-\sigma ,i}=\frac{[p_i-C_{i-\sigma ,i}][p_{i-\sigma}
-C_{i-\sigma ,i}]}{[1-\sum_{k=i-\sigma}^i p_k + C_{i-\sigma ,i}]}e^{-v_c},
\end{equation}
with physically relevant solutions,
\begin{equation}
\label{eq:cij-contact}
C_{i-\sigma,i}=
\frac
{A_i-\left[A_i^2-4e^{-v_c}(e^{-v_c}-1)p_{i-\sigma}p_i\right]^{1/2}}
{2(e^{-v_c}-1)}\,,
\end{equation}
where
\begin{equation}
\label{eq:a-factor}
A_i=1+e^{-v_c}(p_{i-\sigma}+p_i)-\sum_{k=i}^{i-\sigma}p_k\,.
\end{equation}

\begin{figure}[t!]
\includegraphics[scale=0.28]{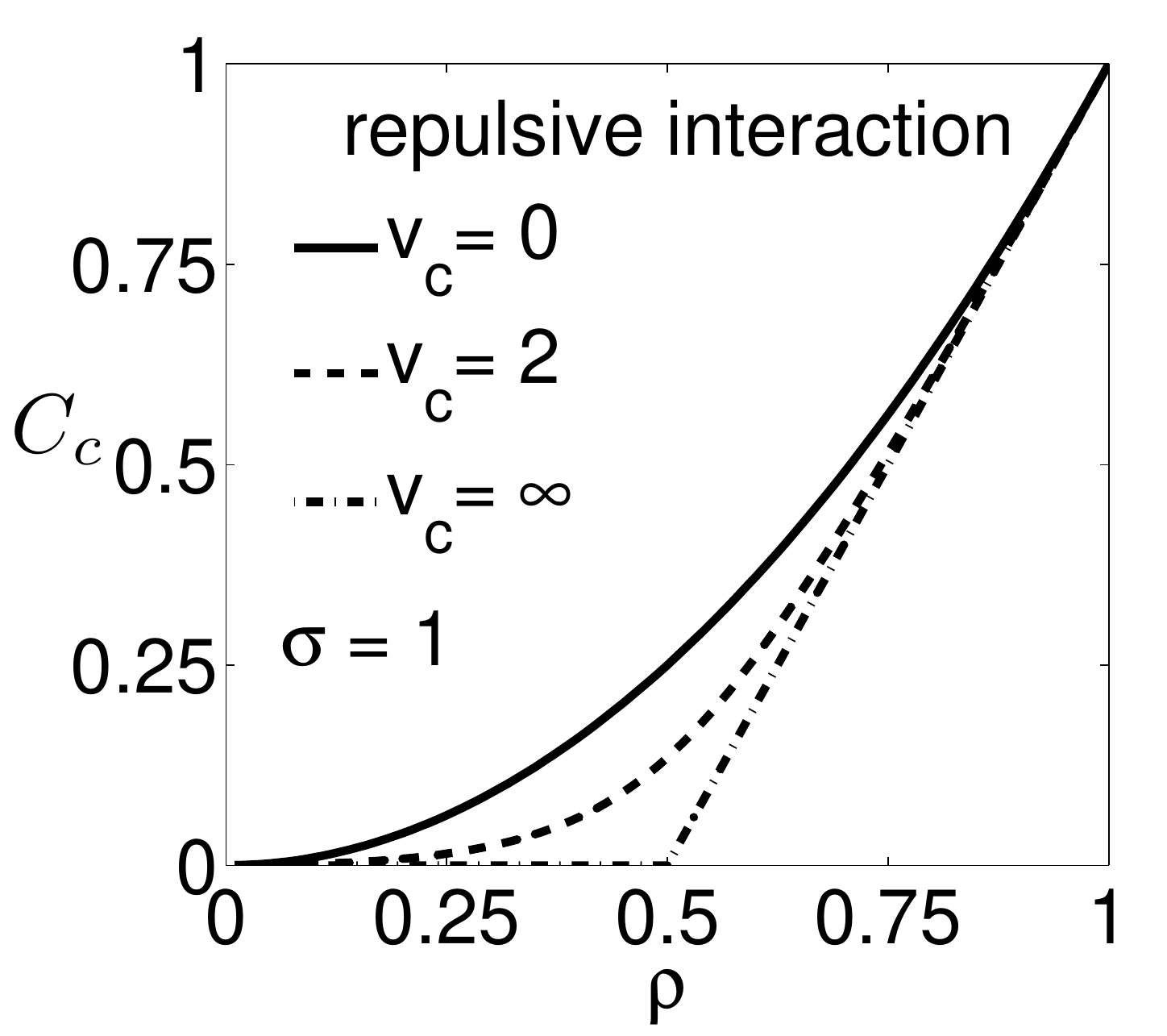}
\includegraphics[scale=0.28]{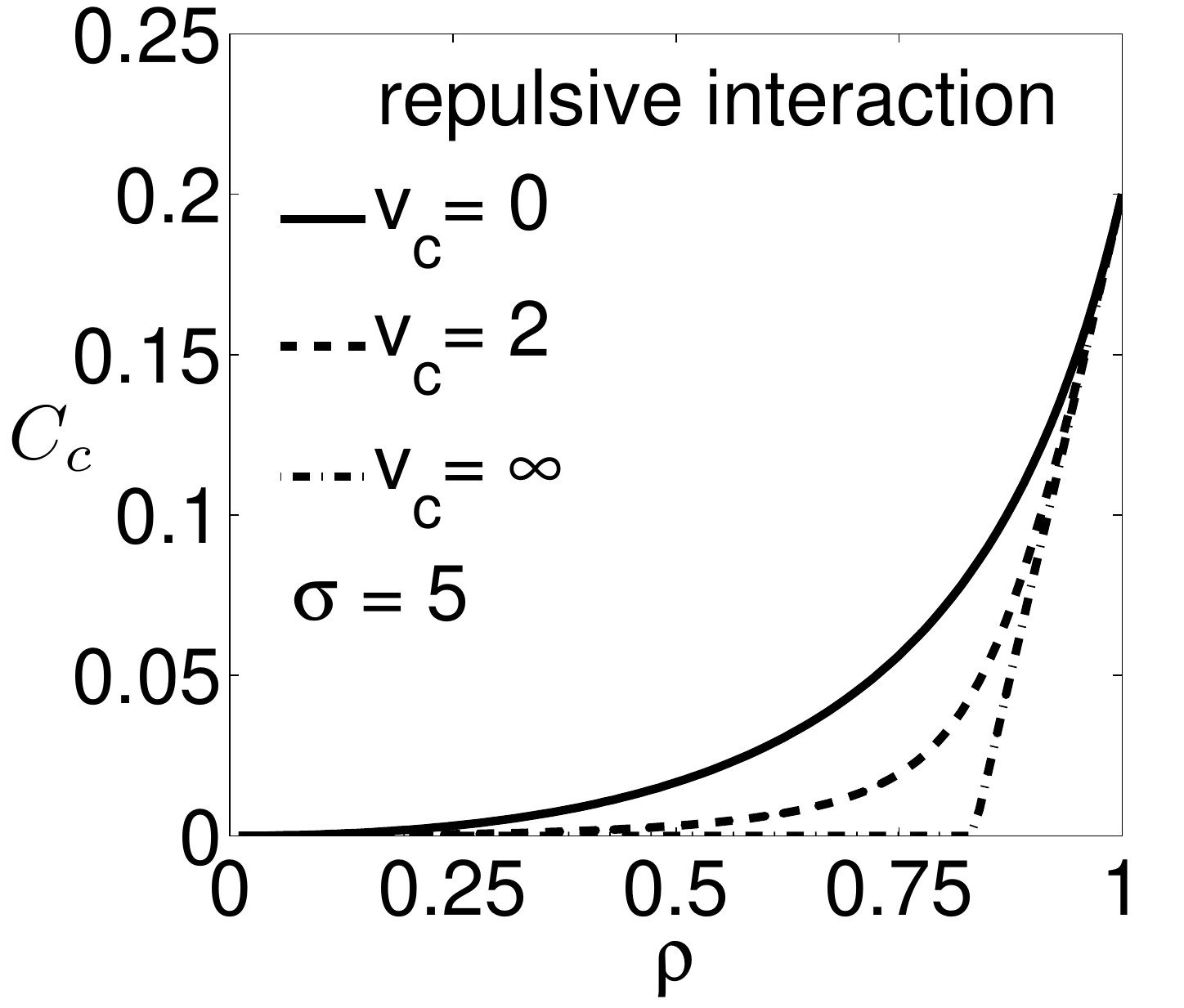}\\
\includegraphics[scale=0.288]{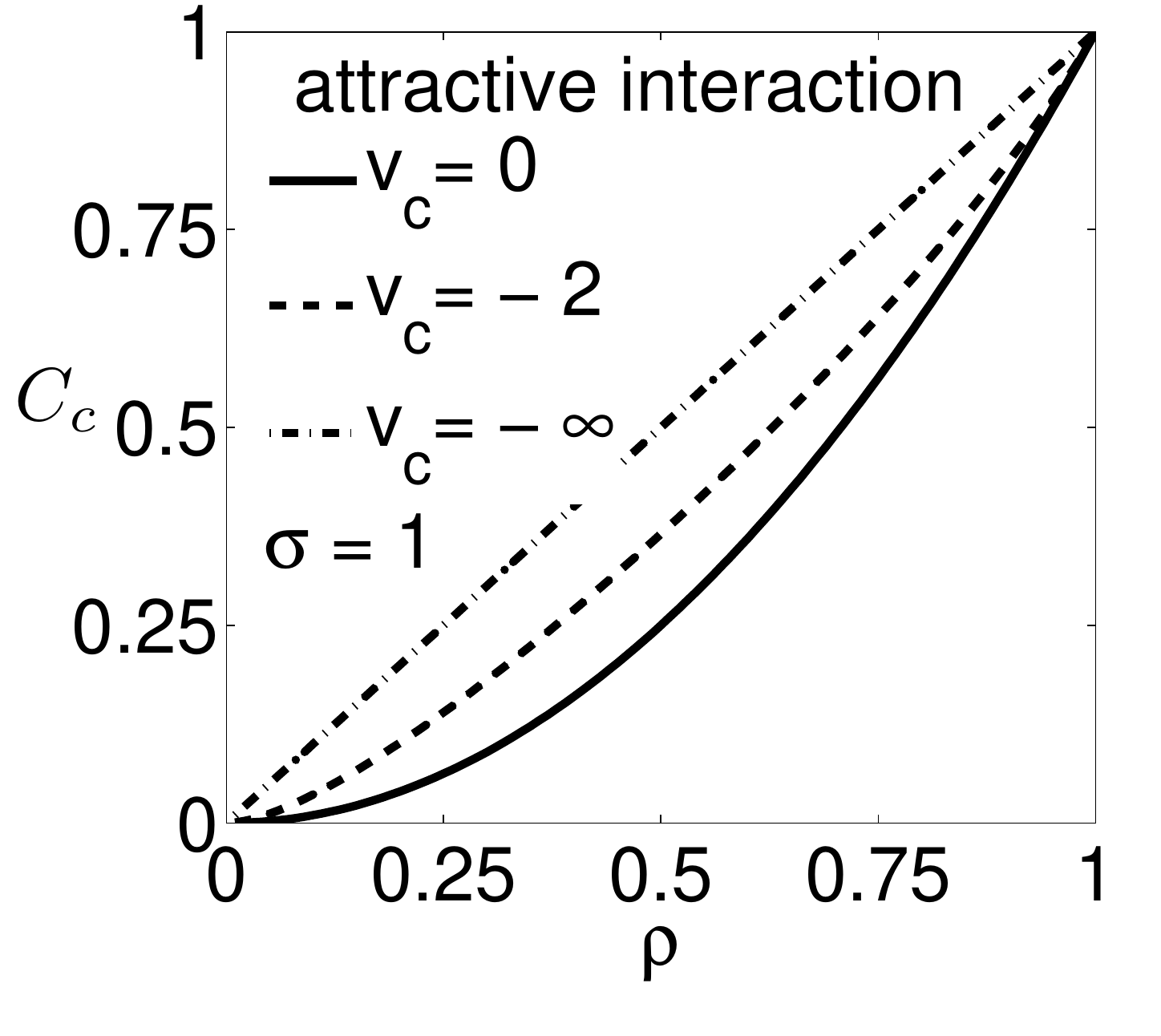}
\includegraphics[scale=0.28]{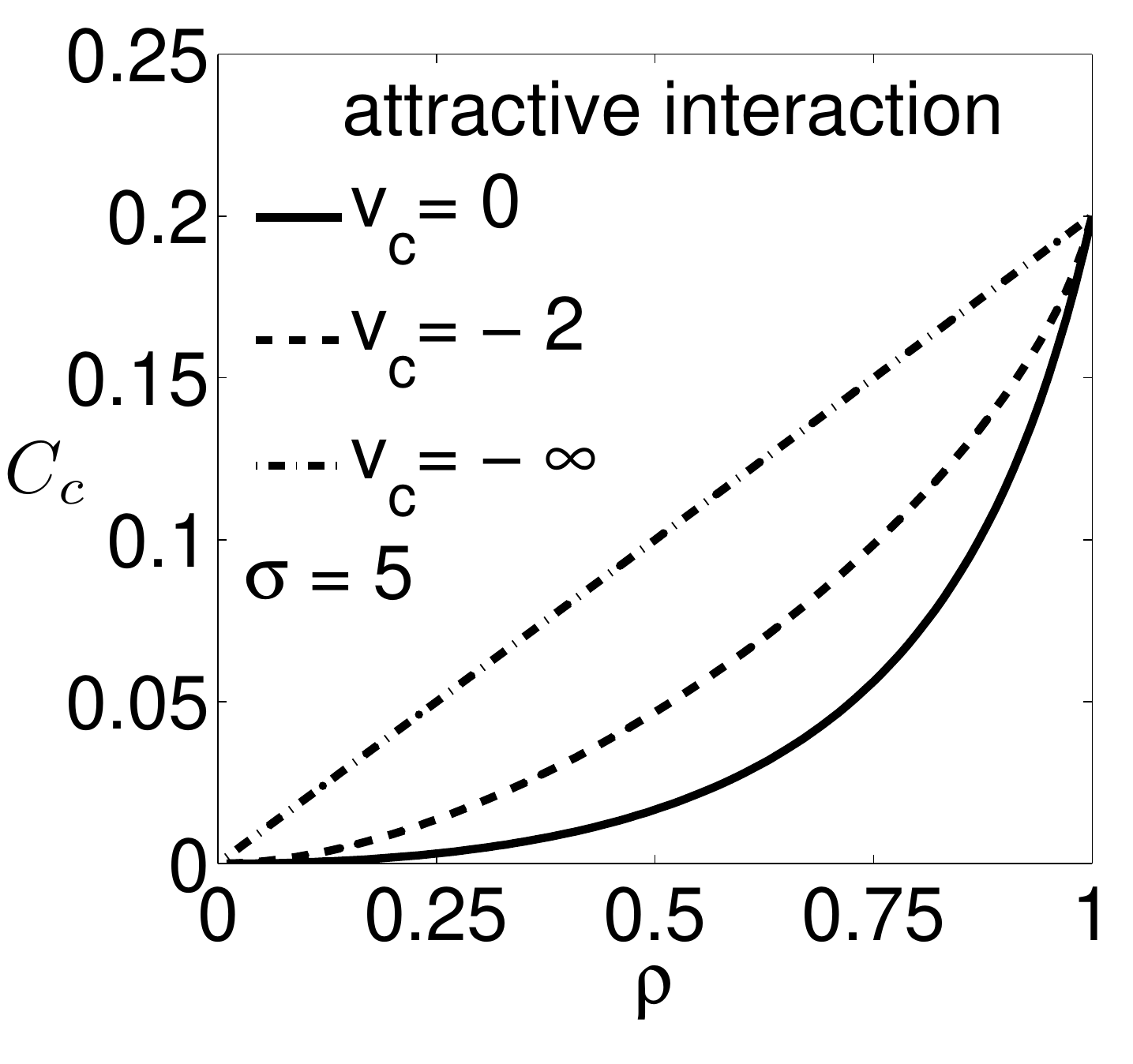}
\caption{\label{fig:prob-contact} Contact correlators
  $C_c=C_{i-\sigma,i}$ for spatially homogeneous systems of hard rods
  of sizes $\sigma=1$ and 5, and different strength of repulsive and
  attractive contact interactions $v_c=v_{i-\sigma,i}$.}
\end{figure}

Figure~\ref{fig:prob-contact} shows, as an example, the contact
correlators $C_c\equiv C_{i-\sigma,i}$ for a spatially homogeneous
bulk system ($u_i=0$) with mean occupation numbers $p_i=p$. For all
graphs shown in the following we use the coverage $\rho=p\sigma$ as
independent variable, which can be interpreted as mass density of
sorts.  We consider rods of sizes $\sigma=1$ and $\sigma=5$ and
contact interactions of zero, finite, and infinite strength
(attractive and repulsive).

The solid curve in each panel represents the result for the
non-interacting case, where $C_c=p^2=\rho^2$ for $\sigma=1$, because
correlations are absent in the simple Fermi lattice gas.  For larger
rod size, we have $C_c>\rho^2/\sigma^2$ for $0<\rho<1$. This is a
consequence of the well-known entropy effect in systems with athermal
exclusion interactions: Bringing neighboring rods closer to each other
gives the remaining rods more configurational freedom and in total the
system more configurational space.  Repulsive interactions $(v_c>0)$
cause dispersal of rods, which suppresses contact correlations.
Attractive interactions $(v_c<0)$, on the other hand, lead to
clustering of rods, which enhances contact correlations.  Infinitely
strong attraction produces a single cluster that grows with $\rho$,
whereas infinitely strong repulsion makes the rods avoid all contacts
if possible, i.e.\ for $\rho<\sigma/(\sigma+1)$. These attributes
account for the (piecewise) linear dependence of $C_c$ on $\rho$.

\section{Density functionals}
\label{sec:densityfunc}

Based on the Gibbs-Bogoliubov inequality the following functional
is defined in density functional theory,
\begin{align}
\Omega[\mathbf{p}]&=\sum_{\mathbf{n}}\chi_{\mathbf{p}}(\mathbf{n})
\left[\ln \chi_{\mathbf{p}}(\mathbf{n})+V(\mathbf{n})+U(\mathbf{n})-
\mu N\right]\nonumber\\
&=F[\mathbf{p}]+
\sum_{k=1}^L (u_k-\mu)p_k\,,
\label{eq:omega}
\end{align}
where $F[\mathbf{p}]=\sum_{\mathbf{n}}
\chi_{\mathbf{p}}(\mathbf{n})[\ln
\chi_{\mathbf{p}}(\mathbf{n})+V(\mathbf{n})]$ is the free energy
functional and the $u_k$ is the external in
Eq.~(\ref{eq:hamiltonian}).  Minimizing $\Omega[\mathbf{p}]$ yields
the equilibrium density profile $\mathbf{p}^{eq}=\{p_i^{eq}\}$.

Substituting the results of Sec.~\ref{sec:microstates} and using the
abbreviated notation, $\Phi(x):=x\ln x$, we can write the free-energy
functional in the form
\begin{align}
F[\mathbf{p}]&=
\sum_{i,j} v_{i,j}C_{i,j}\nonumber\\
&{}+\sum_{s=1}^{L}\Biggl\{
\Phi\Bigl(p_s\!-\!\sum_{i=s-\xi}^{s-\sigma}C_{s,i}\Bigr)\nonumber\\
&\phantom{{}+\sum_{s=1}^{L}\Bigl\{}
{}+\Phi\Bigl(1\!-\!\sum_{i=s-\xi}^sp_i 
\!+\!\sum_{i=s-\xi+\sigma}^s\sum_{j=s-\xi}^{i-\sigma}C_{i,j}\Bigr)\nonumber\\
&\phantom{{}+\sum_{s=1}^{L}\Bigl\{}
{}-\Phi\Bigl(1\!-\!\sum_{i=s-\xi}^{s-1}p_i 
+\sum_{i=s-\xi+\sigma}^{s-1}\sum_{j=s-\xi}^{i-\sigma-1}C_{i,j}\Bigr)\nonumber\\
&\phantom{{}+\sum_{s=1}^{L}\Bigl\{}
{}+\sum_{i=s-\xi}^{s-\sigma}\Bigl\{
\Phi\Bigl(C_{s,i}\Bigr)+\Phi\Bigl(p_{i}-\sum_{j=i+\sigma}^{s}C_{j,i}\Bigr)\nonumber\\
&\phantom{{}+\sum_{s=1}^{L}\Bigl\{{}+\sum_{i=s-\xi}^{s-\sigma}\Bigl\{}
{}-\Phi\Bigl(p_{i}-\sum_{j=i+\sigma}^{s-1}C_{j,i}\Bigr)\Bigr\}\Biggr\}\,,
\label{eq:f}
\end{align}
with the $C_{i,j}=C_{i,j}[\mathbf{p}]$ extracted from Eq.~(\ref{eq:c-p-rel}).
For contact interactions, $v_c\equiv v_{i,i+\sigma}$, it can be
rendered more compactly:
\begin{align}
\label{eq:f-contact-1}
F[\mathbf{p}] &= \sum\limits_s \Bigl\{ C_{s-\sigma ,s}v_{s-\sigma, s} 
+ \Phi\Bigl(p_s-C_{s-\sigma ,s}\Bigr)\\
&{}+\Phi\Bigl(1-\sum_{k=s-\sigma}^s p_k  + C_{s-\sigma ,s}\Bigr)
-\Phi\Bigl(1-\sum_{k=s-\sigma}^{s-1} p_k\Bigr) \nonumber\\
&{}+ \Phi\Bigl(C_{s-\sigma ,s}\Bigr) + 
\Phi\Bigl(p_{s-\sigma}-C_{s-\sigma ,s}\Bigr)-
\Phi\Bigl(p_{s-\sigma}\Bigr)\Bigl\}
\nonumber
\end{align}

For the special case $\sigma=1$ (Ising lattice gas), this functional
agrees with a previous result of Ref.~\onlinecite{Buschle/etal:2000a},
for which Lafuente and Cuesta \cite{Lafuente/Cuesta:2005} derived a
fundamental measure form also.  As it happens, the fundamental measure
form can be extended to hard rods ($\sigma>1$) with contact
interaction by eliminating $v_{i,i+\sigma}C_{i,i+\sigma}$ in
Eq.~(\ref{eq:f-contact-1}) in favor of correlators and densities via
Eq.~(\ref{eq:c-p-rel}):
\begin{align}
\label{eq:f-contact}
F[\mathbf{p}]&=\sum_{s=1}^L\left(\mathcal{F}_{2}[p_{s-\sigma},\ldots,p_s]
-\mathcal{F}_1[p_{s-\sigma},\ldots,p_{s-1}]\right)
\end{align}
where
\begin{align}
\label{eq:f1}
&\mathcal{F}_1[p_{s-\sigma},\ldots,p_{s-1}]=\nonumber\\
&\hspace{2em}p_{s-\sigma}\ln p_{s-\sigma}+(1-\sum_{i=s-\sigma}^{s-1}p_i)
\ln(1 -\sum_{i=s-\sigma}^{s-1}p_i)
\end{align}
and
\begin{align}
\label{eq:f2}
&\mathcal{F}_{2}[p_{s-\sigma},\ldots,p_s]=\nonumber\\
&\hspace{3em}
p_s\ln(p_s-C_{s-\sigma,s})+p_{s-\sigma}\ln(p_{s-\sigma}-C_{s-\sigma,s})\nonumber\\
&\hspace{3em}
{}+(1-\sum_{i=s-\sigma}^sp_i)\ln(1-\sum_{i=s-\sigma}^sp_i + C_{s-\sigma,s})
\end{align}
are the free energy functionals of one-particle and two-particle
cavities, respectively. A one-particle cavity refers to a range of
successive lattice sites, where at most one occupation number can be
one, in analogy to the zero-dimensional cavity in Rosenfeld's
fundamental measure theory in continuous space. For discrete lattice
gas systems, following Lafuente and Cuesta,
\cite{Lafuente/Cuesta:2005} an $m$-particle cavity refers to a range
of successive lattice sites, where at most $m$ occupation numbers can
be one. Notice that the size of an $m$-particle cavity can vary
between $m\sigma$ (minimal $m$-particle cavity) and $(m+1)\sigma-1$
(maximal $m$-particle cavity). In this respect, $\mathcal{F}_1$ in
Eq.~(\ref{eq:f1}) refers to a maximal one-particle cavity and
$\mathcal{F}_2$ in Eq.~(\ref{eq:f2}) to a minimal two-particle cavity.
Fundamental measure forms allow for a straighforward extension to
approximate functionals in higher dimensions that become exact under
dimensional reduction.\cite{Rosenfeld:1989, Lafuente/Cuesta:2005}

\section{Thermodynamics of homogeneous systems}

\begin{figure}[b!]
\includegraphics[scale=0.28]{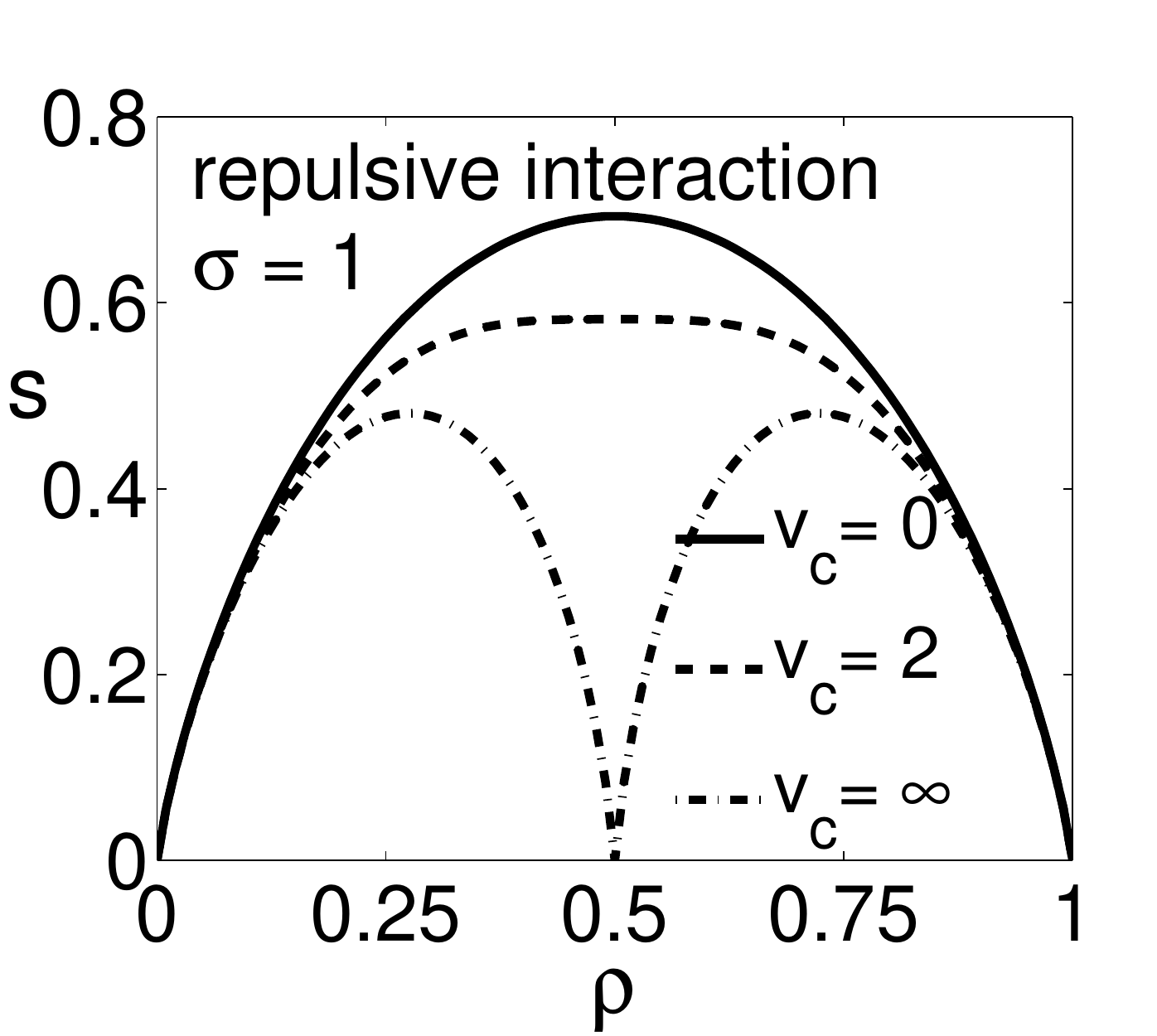}
\includegraphics[scale=0.28]{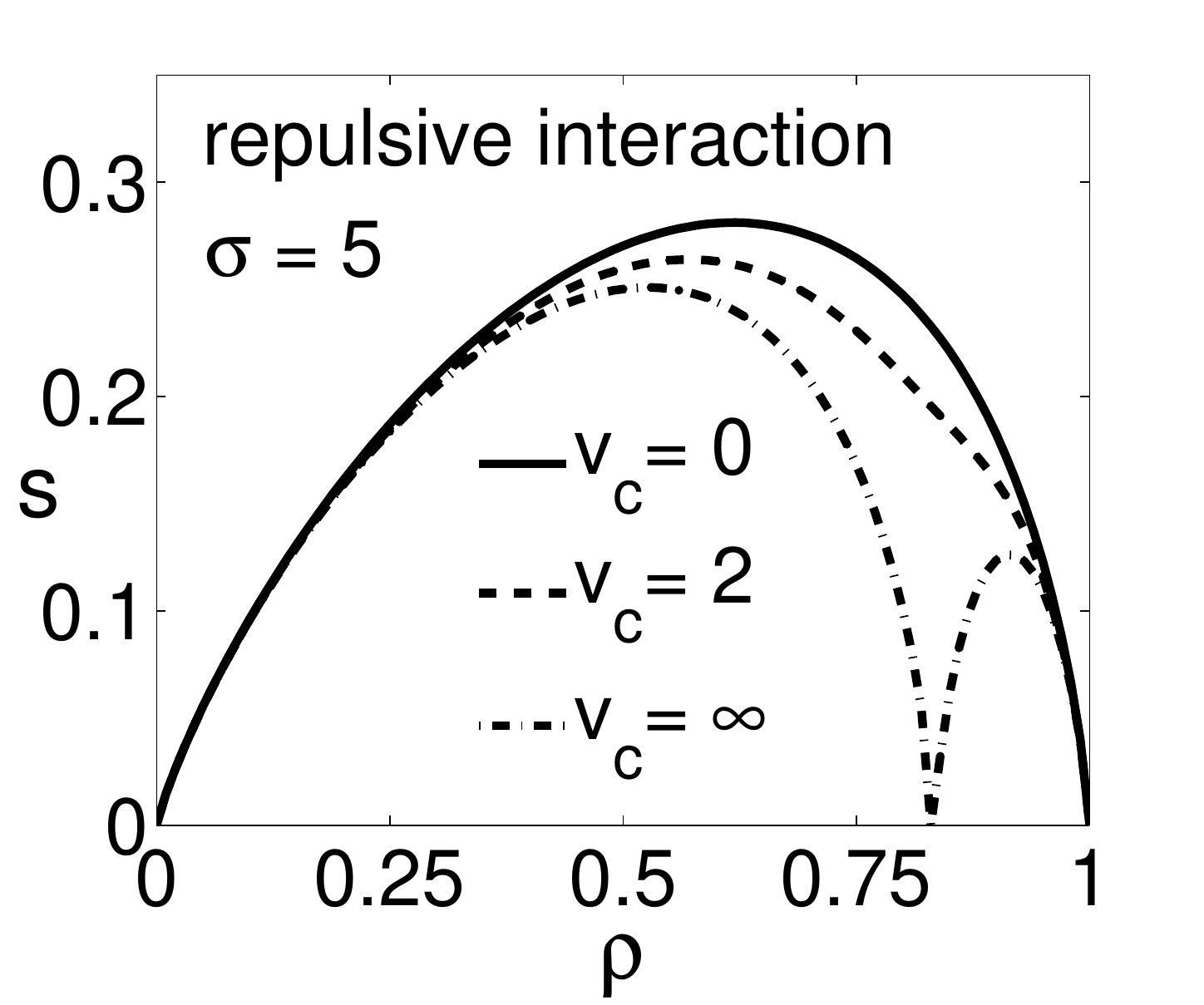}\\
\includegraphics[scale=0.28]{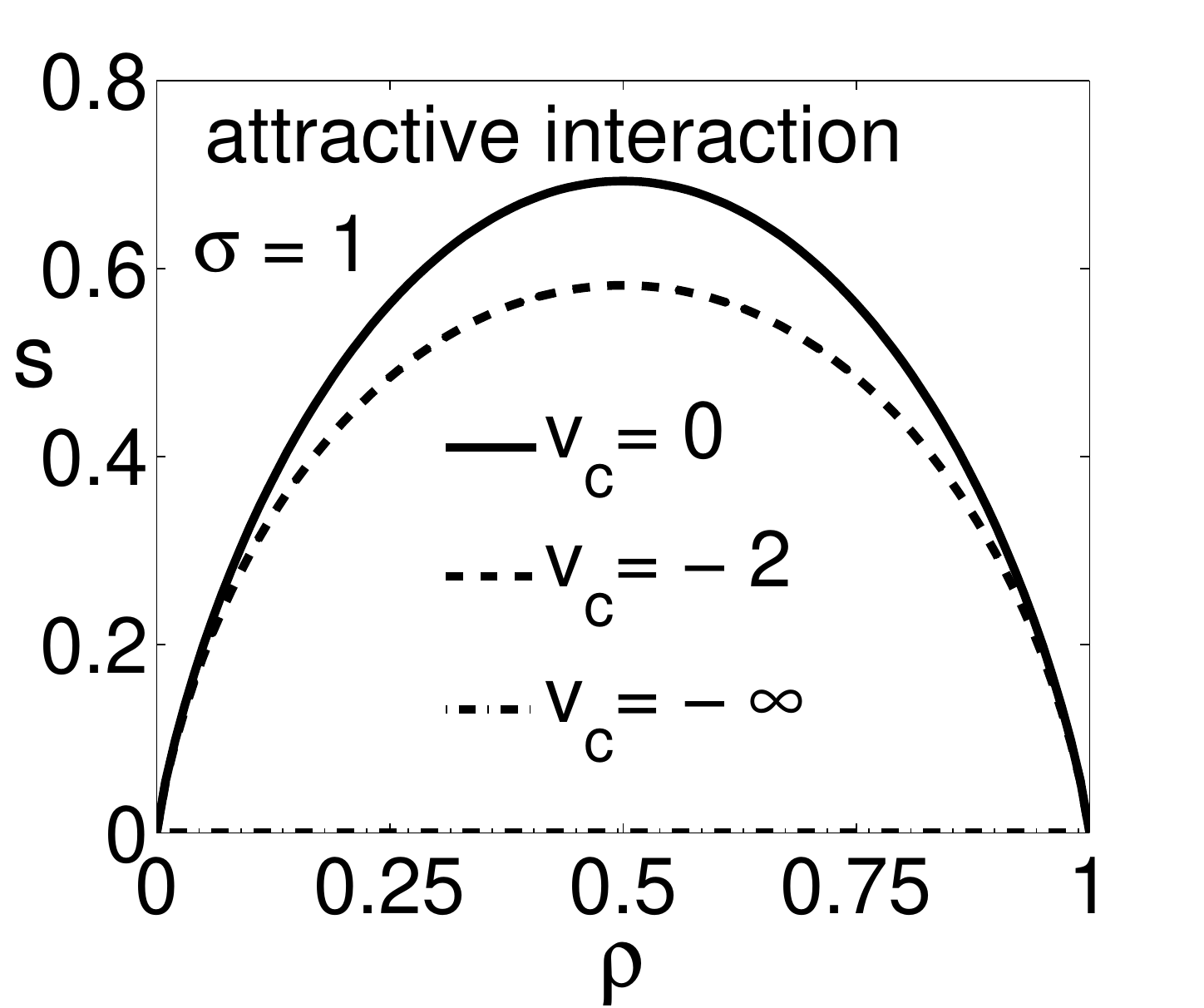}
\includegraphics[scale=0.28]{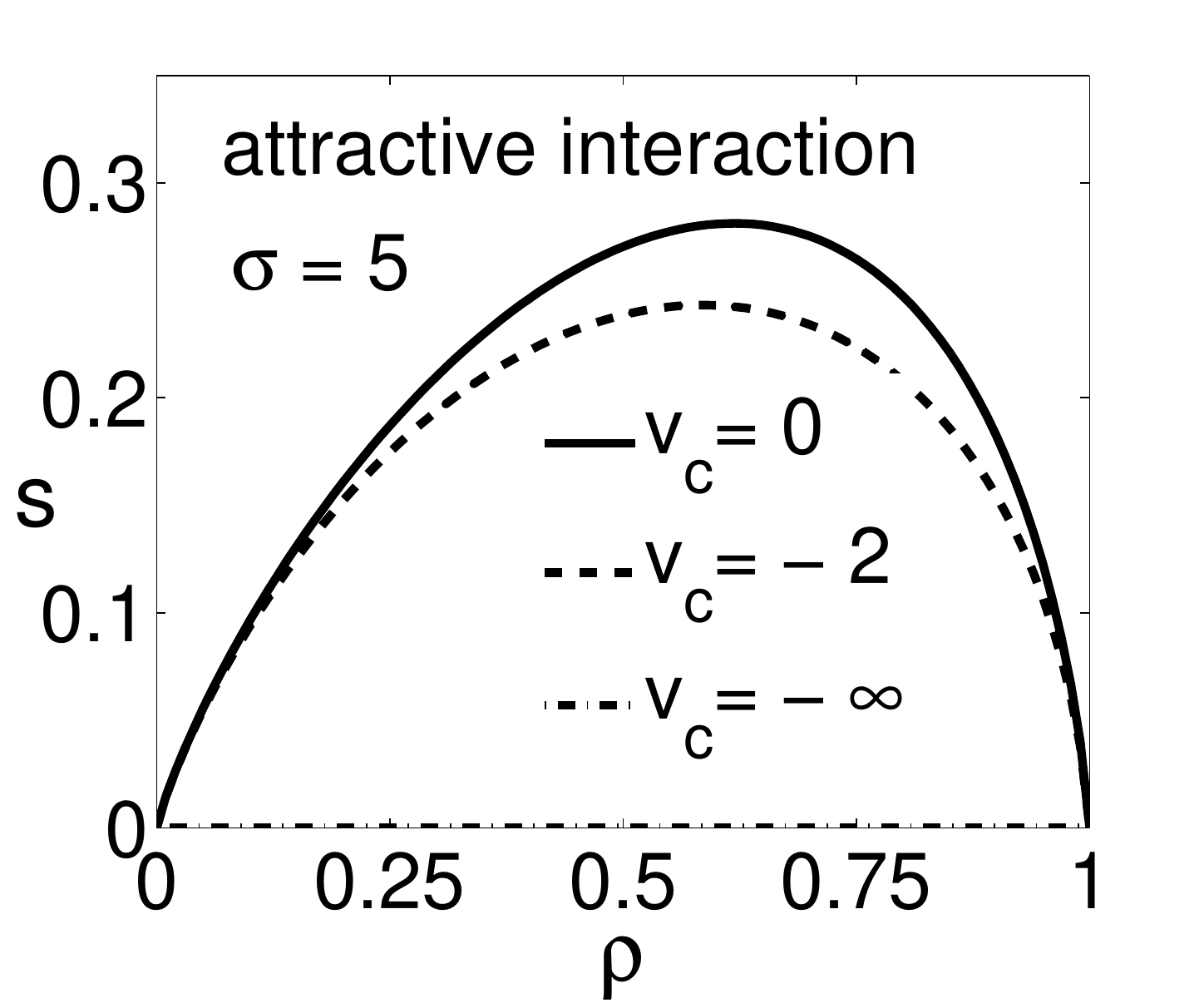}
\caption{\label{fig:entropy} Entropy $s$ per lattice site as a
  function of the coverage $\rho$ for a homogeneous system of hard
  rods of sizes $\sigma=1$ and 5, and various contact interactions
  $v_c$.}
\end{figure}

\begin{figure}[t!]
\includegraphics[scale=0.28]{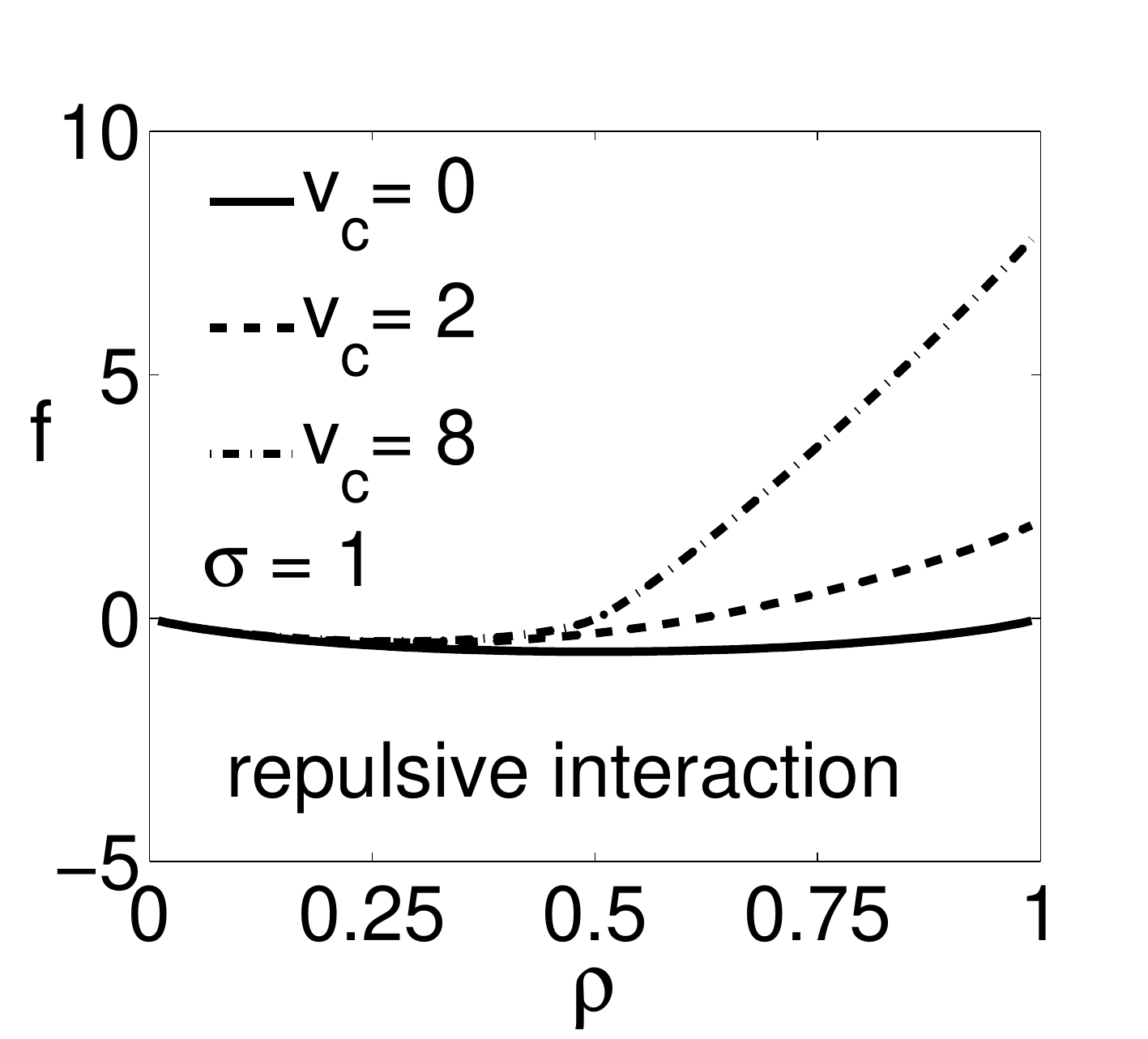}
\includegraphics[scale=0.28]{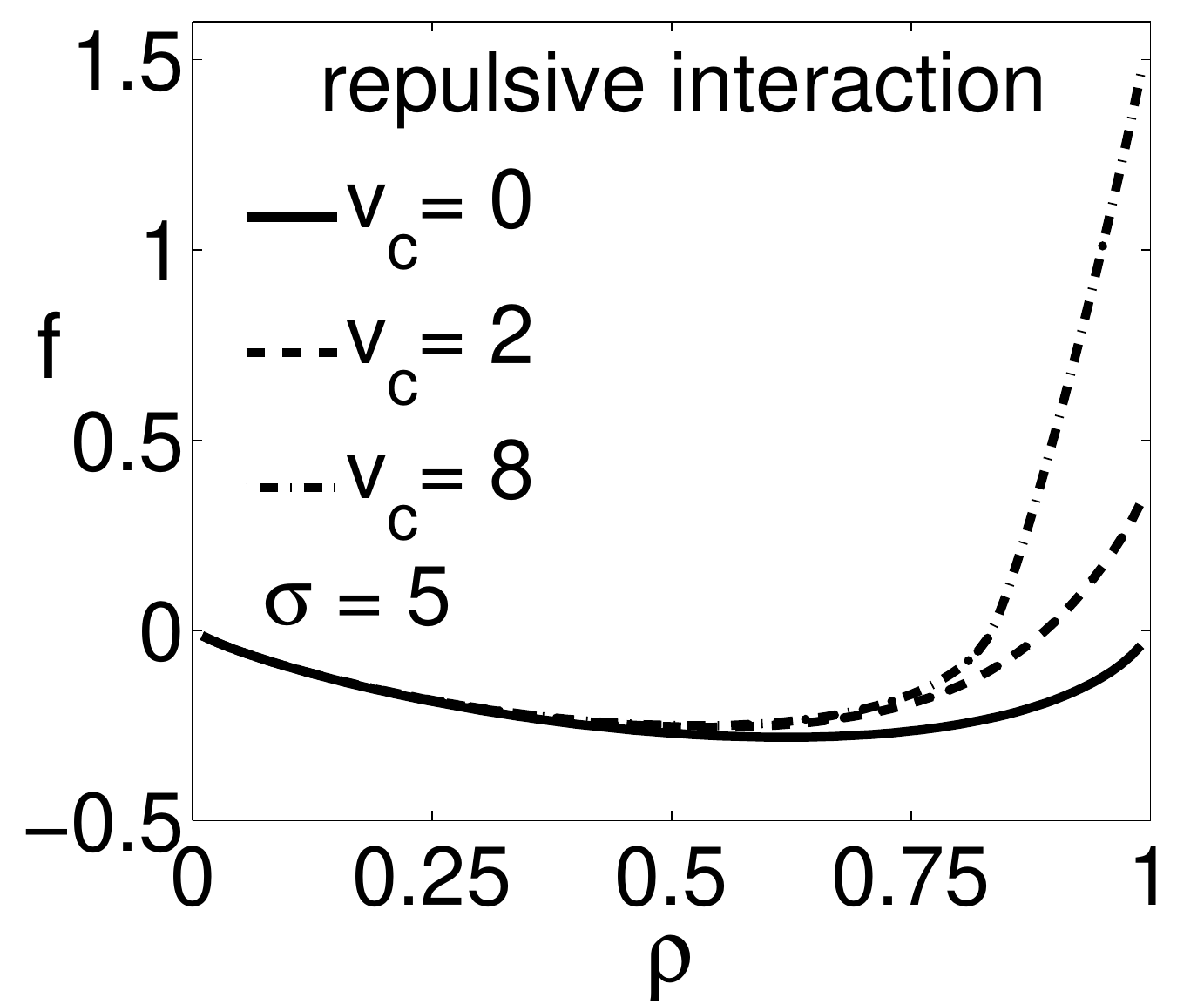}\\
\includegraphics[scale=0.28]{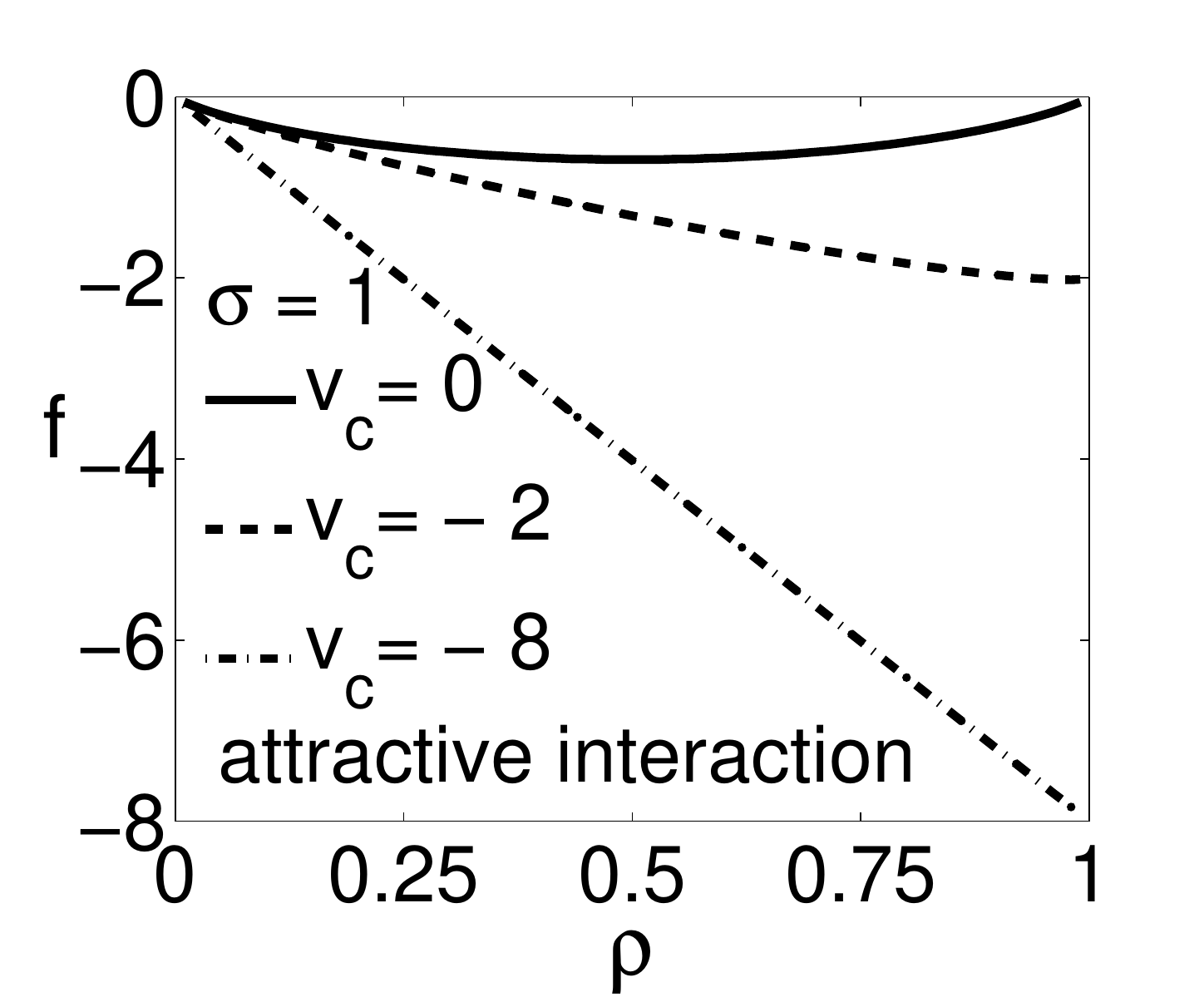}
\includegraphics[scale=0.28]{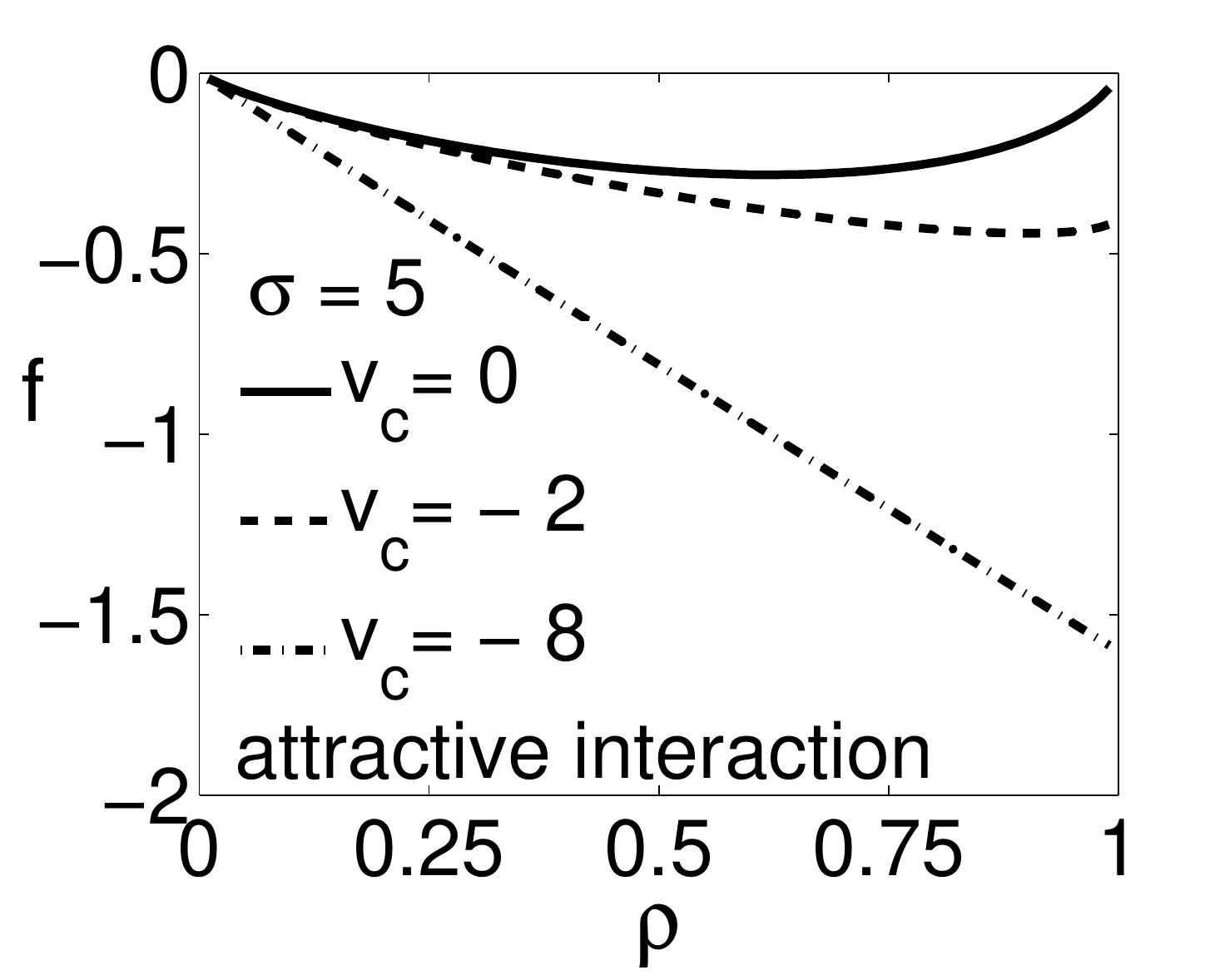}
\caption{\label{fig:free_energy} Free energy per site versus $\rho$
  for a homogeneous system of hard rods of sizes $\sigma=1$ and 5, and
  various contact interactions $v_c$.}
\end{figure}

Here the focus is entirely on bulk thermodynamic properties of
homogeneous systems with attractive or repulsive contact interactions.
The internal energy per site $u_{\scriptscriptstyle\rm
  int}=U_{\scriptscriptstyle\rm int}/L$ is directly related to the
contact correlator $C_c$ from Eq.~(\ref{eq:cij-contact}):
\begin{equation}\label{eq:int-ene} 
  u_{\scriptscriptstyle\rm int}(p)=v_cC_c(p).
\end{equation}
Its dependence on coverage $\rho$ for $\sigma=1,5$ is represented by
the curves in Fig.~\ref{fig:prob-contact}, appropriately rescaled by
$v_c$ (positive or negative).  Not surprisingly, the magnitude of
$u_{\scriptscriptstyle\rm int}$ increases with crowding. We infer the
free energy per site $f=F/L$ from
Eq.~(\ref{eq:f-contact-1}):
\begin{align}
f(p)&=(1-(\sigma +1)p)\ln(1-(\sigma +1)p + C)\nonumber\\
 &+2p\ln(p-C) - (1-\sigma p)\ln(1-\sigma p) - p\ln p.
\end{align}
The entropy per site $s=S/L$ follows directly:
\begin{equation}\label{eq:ent-p} 
s(p)=u(p)-f(p).
\end{equation}
Their dependences on $\rho$ are shown in Figs.~\ref{fig:entropy} and
\ref{fig:free_energy}.

Attractive and repulsive contact interactions both lead to an entropy
reduction.  The underlying causes are different.  For $v_c<0$ the rods
have a tendency to form clusters.  This ordering tendency is largely
independent of coverage, producing a relative entropy reduction that
depends only weakly on $\rho$.  In the limit $v_c\to-\infty$, the rods
form a single cluster, implying $s=0$ for any $\rho$.

For $v_c>0$, by contrast, the rods have a tendency to disperse, i.e.\
to avoid contact.  The associated ordering tendency in the face of
space constraints strongly depends on coverage.  Above a critical
strength $v_c^\star\cong1.95$ of the contact interaction, the entropy develops
a double-hump structure with a minimum at $\rho=1/2$ for $\sigma=1$,
and close to $\sigma/(\sigma+1)$ for $\sigma>1$. At the coverage
$\rho=\sigma/(\sigma+1)$, perfect ordering, corresponding to a
vanishing entropy, is obtained in the limit $v_c\to+\infty$ ($v_c\gg
v_c^\star$). For lower or higher coverages, residual entropy $s$
persists at significant levels.  Entropy profiles similar to those in
Fig.~\ref{fig:entropy} were recently found by a study of jammed
granular matter in a narrow channel.\cite{Gundlach/etal:2013}

A point of some interest is the location of the entropy maximum in
dependence of $\sigma$ and $v_c$ (first maximum for $v_c>v_c^\star$ in
the case of repulsive interactions). The curves show that for $v_c=0$
the coverage of maximum entropy shifts to right from $\rho=1/2$ as the
size of the rods grows from $\sigma=1$ and that, for given $\sigma>1$,
it shift to left as $|v_c|$ increases. This shift can be understood
intuitively by viewing rods and vacancies as two species with numbers
$N_{\rm rod}=pL=\rho L/\sigma$ and $N_{\rm vac} =(1-\rho)L$,
respectively. Neglecting entropy related correlation effects in the
non-interacting case, the number $\sim e^{sL}$ of possible
configurations could then be estimated by $(N_{\rm rod}+N_{\rm
  vac})!/(N_{\rm rod}!N_{\rm vac}!)$. For $\sigma=1$, this is clearly
maximal for equal number of particles and vacancies $(\rho=1/2)$,
while for $\sigma >1$, one has to take into account that $(N_{\rm
  rod}+N_{\rm vac})$ decreases with increasing $\rho$. Accordingly,
the location of the entropy maximum occurs left to
$\rho=\sigma/(\sigma+1)$, where $N_{\rm rod}=N_{\rm vac}$. In the
interacting case, the coverage of maximum entropy moves to the left
for increasing interaction strength, because the enhanced
configurational restrictions for larger $|v_c|$ can be partly
compensated by lowering $\rho$. Analytical results for the entropy of
hard rods with nearest-neighbor interactions in one-dimensional
continuum have been earlier obtained by Percus.\cite{Percus:1989} Its
overall behavior as function of the density in the case of contact
interaction agrees with the one displayed in Fig.~\ref{fig:entropy}.

Because of absence of phase transitions in one dimension (if not
considering ``exotic'' cases of interaction with particular long-range
behavior \cite{Berlin/Kac:1952}) the free energy shown in
Fig.~\ref{fig:free_energy} does not show any peculiarities as a
function of $\rho$. In the case of attractive interaction, it
approaches the line $f\sim -|v_c|\rho/\sigma$ for large $|v_c|$ due to
the aggregation of the rods into one cluster. For strong repulsive
interactions, i.e.\ for $v_c$ significantly larger than $v_c^\star$,
the free energy essentially follows the (negative) entropy for
$\rho\lesssim\sigma/(\sigma+1)$, increases linearly for
$\rho\gtrsim\sigma/(\sigma+1)$ due to the linear increase of
non-avoidable contacts until reaching $C_cv_c\simeq v_c/\sigma$ for
$\rho\to1$.

\begin{figure}[t!]
\includegraphics[scale=0.285]{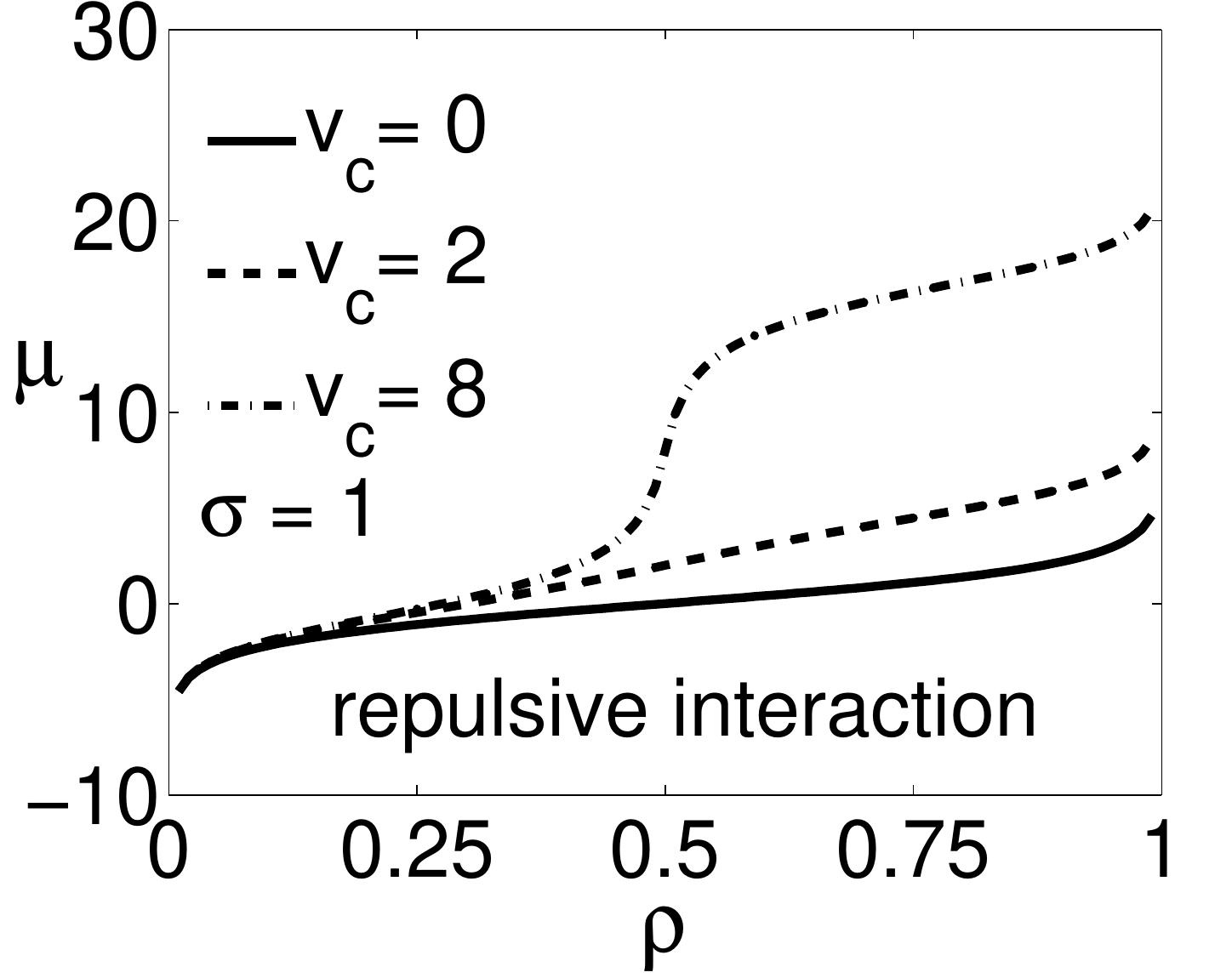}
\includegraphics[scale=0.285]{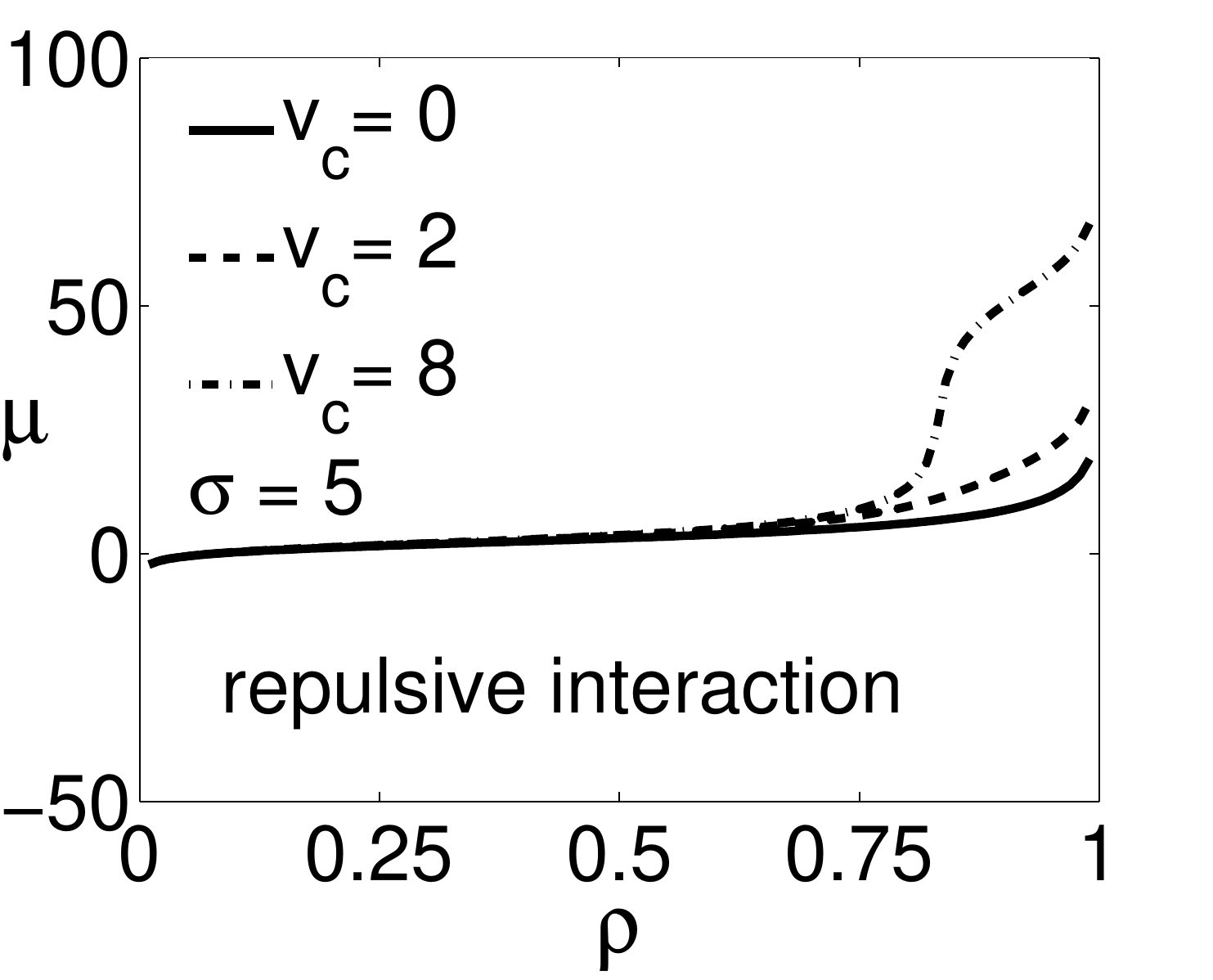}\\
\includegraphics[scale=0.285]{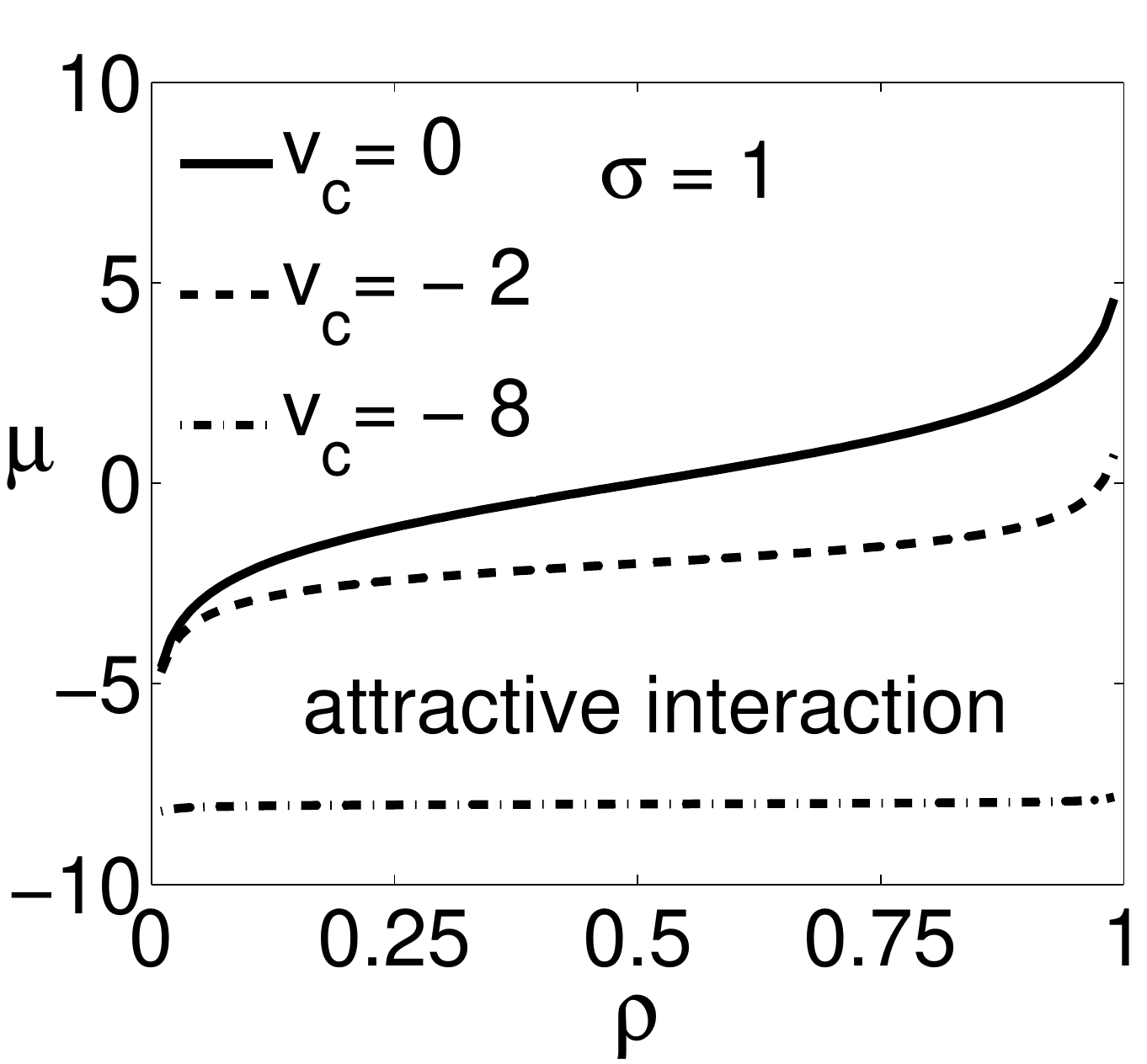}
\includegraphics[scale=0.285]{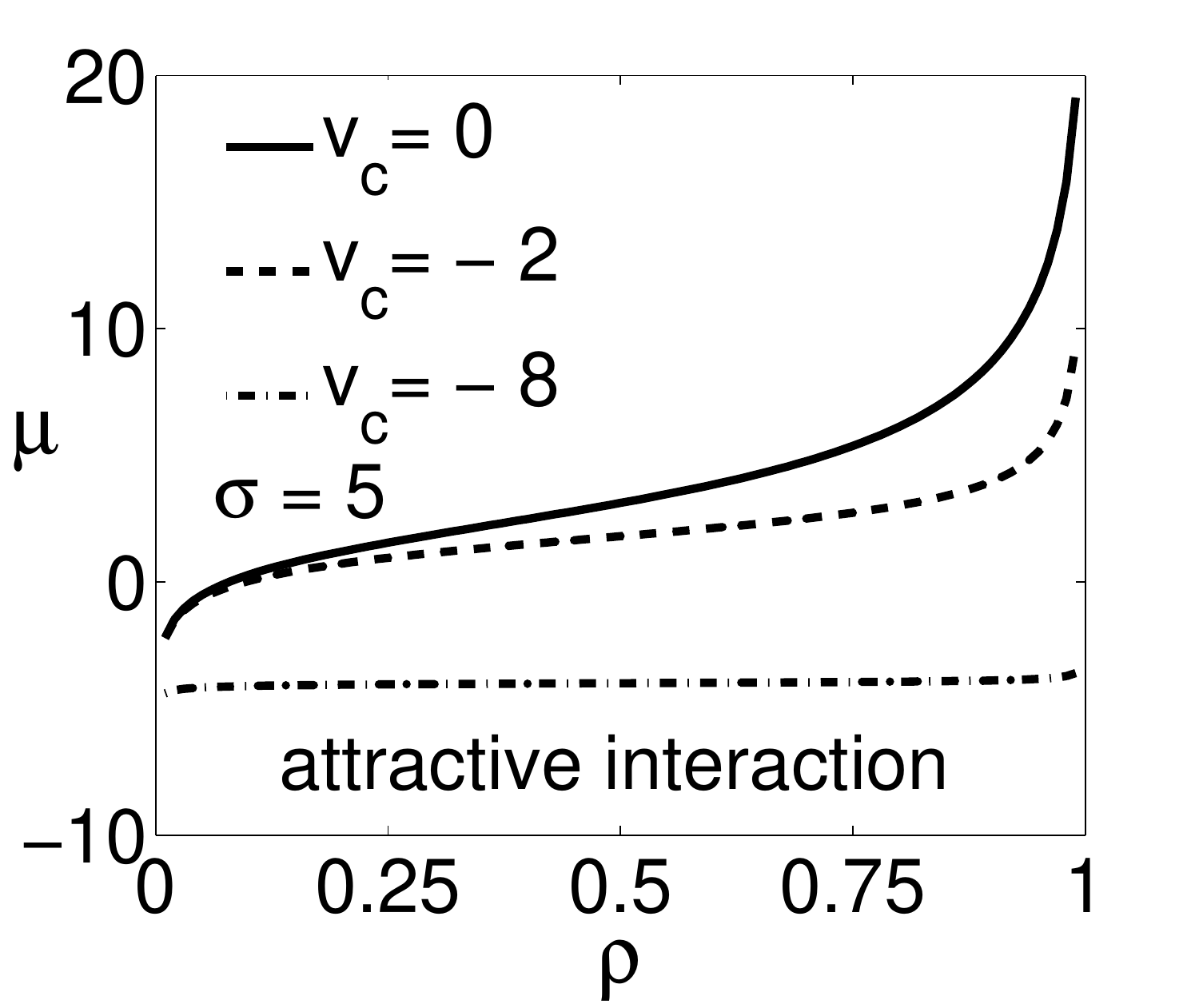}
\caption{\label{fig:fugacity} Chemical potential $\mu$ as a function
  of the coverage $\rho$ for a homogeneous system of hard rods of
  sizes $\sigma=1$ and 5, and various contact interactions $v_c$.}
\end{figure}

For repulsive interactions, the density $\rho=\sigma/(\sigma+1)$
should also show up as a particular value in the behavior of the
chemical potential, because the free energy amount to add a rod to the
system is expected to increase strongly around this point. The
chemical potential $\mu=\partial f/\partial\rho$ is given by
\begin{equation}
\mu= \ln\left\{\frac{\sigma C}{\rho e^{-v}e^{1-\sigma}}\right\}
+\sigma\ln\left\{
\left[\frac{1-\rho}{1-(\sigma +1)\rho/\sigma +C}\right]\right\}
\end{equation}
and plotted in Fig.~\ref{fig:fugacity} as function of $\rho$. It
indeed shows a step-like change around $\rho=\sigma/(\sigma+1)$ for
strong repulsive interactions, which could be utilized for determining
the rod size from thermodynamic measurements. For comparison the
behavior for attractive interaction is also displayed in
Fig.~\ref{fig:fugacity}.

\section{Conclusions}

Our solution for the distribution of microstates of hard-rod lattice
gases with a general nearest-neighbor-range interaction potential
provides a promising basis for future studies related to
applications. For example, the results can be utilized to describe
formation of molecular nanowires on surfaces, where molecules interact
via van der Waals interactions and form hydrogen bonds when coming
into contact. In the model, such situation could be accounted for by
an interaction profile $v_{i,j}$ with strong attractive interaction
$v_{i,i+\sigma}<0$ at contact distance $\sigma$ and a smoothly
decreasing, weaker attractive part for $\sigma<|i-j|<2\sigma$. Also,
based on experiences in other contexts, \cite{Dierl/etal:2011,
  Dierl/etal:2012} the explicit exact expressions for the density
functionals should allow one to faithfully study the kinetics of wire
formation by employing time-dependent density functional
theory. Similarly, our results may be helpful in the future to treat
diffusion of molecules through nanopores or membrane channels.

Our derivation of a fundamental measure form of the hard-rod lattice
gas with contact interaction enables a straightforward extension to
higher dimensions. This should be useful to account for transitions
between different phases in corresponding systems, which generally
resemble nematic (and other) phases of liquid
crystals.\cite{Martinez:2007, Martinez/etal:2008, Heras/etal:2010} It
has been shown \cite{Lafuente/Cuesta:2002b, Lafuente/Cuesta:2003a} for
athermal hard-rod lattice gases ($v_{i,j}=0$) that extensions to
higher dimensions are a powerful means to treat corresponding phase
transitions. Up to now, we did not succeed to identify fundamental
measure forms for general nearest-neighbor-range interactions
$v_{i,j}$. However, there seem to be other possibilities of
extensions, which become exact under dimensional reduction. These will
be explored elsewhere.

Considering the core in the derivation of the distribution of
microstates, it is important to realize that the procedure can in
principle be extended to interactions of longer range covering several
rod lengths. For a given range $\xi$, the exclusion constraint leads
to a natural decomposition of the set
$\{\mathbf{n}\}_s=\{n_s,\ldots,n_{s-\xi}\}$ into ranges covering the
lattice sites $s-\sigma+1,\ldots,s$ (first range),
$s-2\sigma+1,\ldots,s-\sigma$ (second range), and so on.  In each of
these ranges at most one occupation number can be equal to one. The
total number of ranges that need to be taken into account is
$\lceil\xi/\sigma\rceil$, where $\lceil x\rceil$ denotes the integer
ceiling division, i.e.\ the smallest integer larger than $x$.
Accordingly, we would need to consider higher-order correlators
$C(1_{i_1},1_{i_2},1_{i_3},\ldots,...)$, where $1_{i_k}$, specifies
the location of the occupied site in the $k$th range as in
Sec.~\ref{subsec:jointprob}. The distribution of microstates can then
be expressed in term of these correlators
$C(1_{i_1},1_{i_2},1_{i_3},\ldots,...)$ and the densities $p_i$.  By
equating with the Boltzmann formula for simple configurations,
relations between the $C(1_{i_1},1_{i_2},1_{i_3},\ldots,...)$ and the
$p_i$ eventually can be obtained, which need to be solved for
expressing the $C(1_{i_1},1_{i_2},1_{i_3},\ldots,...)$ in terms of the
$p_i$.

By working out the long-range asymptotic behavior of correlation
functions it should be possible also to study the occurrence and
behavior of Widom-Fisher lines.\cite{Fisher/Widom:1969} These lines
separate regions in the temperature-density and temperature-pressure
planes, where in one region the pair correlation decays monotonically,
while in the other region it oscillates with decaying amplitude. The
abrupt change in the behavior at the lines takes place without any
singularities in the thermodynamics. In the original work,
\cite{Fisher/Widom:1969} transition lines were calculated for a
one-dimensional continuum fluid with square-well interaction potential
and a lattice fluid as considered in this work, corresponding to hard
rods of size $\sigma=2$ with first-neighbor coupling. Access to the
correlation properties of systems with larger $\sigma$ and
interactions of longer range may allow one to gain deeper insight into
general features of these lines.

Let us finally note that the rod length defines a length scale
independent of the lattice spacing, which implies that a continuum
limit of the results should be accessible. This, the possible
extension to larger interaction ranges, and evaluations of
Widom-Fisher lines open new and challenging possibilities for further
investigations.

\begin{acknowledgements}
  B.~B. would like to thank the Deutsche Akademische Austauschdienst
  (DAAD) for financial support.
\end{acknowledgements}


\begin{thebibliography}{58}%
\makeatletter
\providecommand \@ifxundefined [1]{%
 \@ifx{#1\undefined}
}%
\providecommand \@ifnum [1]{%
 \ifnum #1\expandafter \@firstoftwo
 \else \expandafter \@secondoftwo
 \fi
}%
\providecommand \@ifx [1]{%
 \ifx #1\expandafter \@firstoftwo
 \else \expandafter \@secondoftwo
 \fi
}%
\providecommand \natexlab [1]{#1}%
\providecommand \enquote  [1]{``#1''}%
\providecommand \bibnamefont  [1]{#1}%
\providecommand \bibfnamefont [1]{#1}%
\providecommand \citenamefont [1]{#1}%
\providecommand \href@noop [0]{\@secondoftwo}%
\providecommand \href [0]{\begingroup \@sanitize@url \@href}%
\providecommand \@href[1]{\@@startlink{#1}\@@href}%
\providecommand \@@href[1]{\endgroup#1\@@endlink}%
\providecommand \@sanitize@url [0]{\catcode `\\12\catcode `\$12\catcode
  `\&12\catcode `\#12\catcode `\^12\catcode `\_12\catcode `\%12\relax}%
\providecommand \@@startlink[1]{}%
\providecommand \@@endlink[0]{}%
\providecommand \url  [0]{\begingroup\@sanitize@url \@url }%
\providecommand \@url [1]{\endgroup\@href {#1}{\urlprefix }}%
\providecommand \urlprefix  [0]{URL }%
\providecommand \Eprint [0]{\href }%
\providecommand \doibase [0]{http://dx.doi.org/}%
\providecommand \selectlanguage [0]{\@gobble}%
\providecommand \bibinfo  [0]{\@secondoftwo}%
\providecommand \bibfield  [0]{\@secondoftwo}%
\providecommand \translation [1]{[#1]}%
\providecommand \BibitemOpen [0]{}%
\providecommand \bibitemStop [0]{}%
\providecommand \bibitemNoStop [0]{.\EOS\space}%
\providecommand \EOS [0]{\spacefactor3000\relax}%
\providecommand \BibitemShut  [1]{\csname bibitem#1\endcsname}%
\let\auto@bib@innerbib\@empty
\bibitem [{\citenamefont {Nieswand}\ \emph {et~al.}(1993)\citenamefont
  {Nieswand}, \citenamefont {Dieterich},\ and\ \citenamefont
  {Majhofer}}]{Nieswand/etal:1993a}%
  \BibitemOpen
  \bibfield  {author} {\bibinfo {author} {\bibfnamefont {M.}~\bibnamefont
  {Nieswand}}, \bibinfo {author} {\bibfnamefont {W.}~\bibnamefont {Dieterich}},
  \ and\ \bibinfo {author} {\bibfnamefont {A.}~\bibnamefont {Majhofer}},\
  }\href {\doibase {10.1103/PhysRevE.47.718}} {\bibfield  {journal} {\bibinfo
  {journal} {Phys. Rev. E}\ }\textbf {\bibinfo {volume} {47}},\ \bibinfo
  {pages} {718} (\bibinfo {year} {1993})}\BibitemShut {NoStop}%
\bibitem [{\citenamefont {Reinel}\ and\ \citenamefont
  {Dieterich}(1996)}]{Reinel/Dieterich:1996}%
  \BibitemOpen
  \bibfield  {author} {\bibinfo {author} {\bibfnamefont {D.}~\bibnamefont
  {Reinel}}\ and\ \bibinfo {author} {\bibfnamefont {W.}~\bibnamefont
  {Dieterich}},\ }\href {\doibase {10.1063/1.471150}} {\bibfield  {journal}
  {\bibinfo  {journal} {J. Chem. Phys.}\ }\textbf {\bibinfo {volume} {104}},\
  \bibinfo {pages} {5234} (\bibinfo {year} {1996})}\BibitemShut {NoStop}%
\bibitem [{\citenamefont {Reinhard}\ \emph {et~al.}(2000)\citenamefont
  {Reinhard}, \citenamefont {W.Dieterich}, \citenamefont {Maass},\ and\
  \citenamefont {Frisch}}]{Reinhard/etal:2000}%
  \BibitemOpen
  \bibfield  {author} {\bibinfo {author} {\bibfnamefont {J.}~\bibnamefont
  {Reinhard}}, \bibinfo {author} {\bibnamefont {W.Dieterich}}, \bibinfo
  {author} {\bibfnamefont {P.}~\bibnamefont {Maass}}, \ and\ \bibinfo {author}
  {\bibfnamefont {H.~L.}\ \bibnamefont {Frisch}},\ }\href {\doibase
  10.1103/PhysRevE.61.422} {\bibfield  {journal} {\bibinfo  {journal} {Phys.
  Rev. E}\ }\textbf {\bibinfo {volume} {61}},\ \bibinfo {pages} {422} (\bibinfo
  {year} {2000})}\BibitemShut {NoStop}%
\bibitem [{\citenamefont {Gouyet}\ \emph {et~al.}(2003)\citenamefont {Gouyet},
  \citenamefont {Plapp}, \citenamefont {Dieterich},\ and\ \citenamefont
  {Maass}}]{Gouyet/etal:2003}%
  \BibitemOpen
  \bibfield  {author} {\bibinfo {author} {\bibfnamefont {J.-F.}\ \bibnamefont
  {Gouyet}}, \bibinfo {author} {\bibfnamefont {M.}~\bibnamefont {Plapp}},
  \bibinfo {author} {\bibfnamefont {W.}~\bibnamefont {Dieterich}}, \ and\
  \bibinfo {author} {\bibfnamefont {P.}~\bibnamefont {Maass}},\ }\href@noop {}
  {\bibfield  {journal} {\bibinfo  {journal} {Adv. Phys.}\ }\textbf {\bibinfo
  {volume} {52}},\ \bibinfo {pages} {523} (\bibinfo {year} {2003})}\BibitemShut
  {NoStop}%
\bibitem [{\citenamefont {Cuesta}\ \emph {et~al.}(2005)\citenamefont {Cuesta},
  \citenamefont {Lafuente},\ and\ \citenamefont {Schmidt}}]{Cuesta/etal:2005}%
  \BibitemOpen
  \bibfield  {author} {\bibinfo {author} {\bibfnamefont {J.~A.}\ \bibnamefont
  {Cuesta}}, \bibinfo {author} {\bibfnamefont {L.}~\bibnamefont {Lafuente}}, \
  and\ \bibinfo {author} {\bibfnamefont {M.}~\bibnamefont {Schmidt}},\ }\href
  {\doibase {10.1103/PhysRevE.72.031405}} {\bibfield  {journal} {\bibinfo
  {journal} {{Phys. Rev. E}}\ }\textbf {\bibinfo {volume} {{72}}},\ \bibinfo
  {pages} {{031405}} (\bibinfo {year} {{2005}})}\BibitemShut {NoStop}%
\bibitem [{\citenamefont {Kierlik}\ \emph {et~al.}(2001)\citenamefont
  {Kierlik}, \citenamefont {Monson}, \citenamefont {Rosinderg}, \citenamefont
  {Sarkisov},\ and\ \citenamefont {Tarjus}}]{Kierlik/etal:2001}%
  \BibitemOpen
  \bibfield  {author} {\bibinfo {author} {\bibfnamefont {E.}~\bibnamefont
  {Kierlik}}, \bibinfo {author} {\bibfnamefont {P.~A.}\ \bibnamefont {Monson}},
  \bibinfo {author} {\bibfnamefont {M.~L.}\ \bibnamefont {Rosinderg}}, \bibinfo
  {author} {\bibfnamefont {L.}~\bibnamefont {Sarkisov}}, \ and\ \bibinfo
  {author} {\bibfnamefont {G.}~\bibnamefont {Tarjus}},\ }\href {\doibase
  10.1103/PhysRevLett.87.055701} {\bibfield  {journal} {\bibinfo  {journal}
  {Phys. Rev. Lett.}\ }\textbf {\bibinfo {volume} {87}},\ \bibinfo {pages}
  {055701} (\bibinfo {year} {2001})}\BibitemShut {NoStop}%
\bibitem [{\citenamefont {Azbel}(1979)}]{Azbel:1979}%
  \BibitemOpen
  \bibfield  {author} {\bibinfo {author} {\bibfnamefont {M.~Y.}\ \bibnamefont
  {Azbel}},\ }\href {\doibase 10.1103/PhysRevA.20.1671} {\bibfield  {journal}
  {\bibinfo  {journal} {Phys. Rev. A}\ }\textbf {\bibinfo {volume} {20}},\
  \bibinfo {pages} {1671} (\bibinfo {year} {1979})}\BibitemShut {NoStop}%
\bibitem [{\citenamefont {Rosenfeld}(1989)}]{Rosenfeld:1989}%
  \BibitemOpen
  \bibfield  {author} {\bibinfo {author} {\bibfnamefont {Y.}~\bibnamefont
  {Rosenfeld}},\ }\href {\doibase 10.1103/PhysRevLett.63.980} {\bibfield
  {journal} {\bibinfo  {journal} {Phys. Rev. Lett.}\ }\textbf {\bibinfo
  {volume} {63}},\ \bibinfo {pages} {980} (\bibinfo {year} {1989})}\BibitemShut
  {NoStop}%
\bibitem [{\citenamefont {Lafuente}\ and\ \citenamefont
  {Cuesta}(2002)}]{Lafuente/Cuesta:2002b}%
  \BibitemOpen
  \bibfield  {author} {\bibinfo {author} {\bibfnamefont {L.}~\bibnamefont
  {Lafuente}}\ and\ \bibinfo {author} {\bibfnamefont {J.~A.}\ \bibnamefont
  {Cuesta}},\ }\href {\doibase {10.1088/0953-8984/14/46/314}} {\bibfield
  {journal} {\bibinfo  {journal} {J. Phys.: Condens. Matter}\ }\textbf
  {\bibinfo {volume} {14}},\ \bibinfo {pages} {12079} (\bibinfo {year}
  {2002})}\BibitemShut {NoStop}%
\bibitem [{\citenamefont {Schmidt}\ \emph {et~al.}(2003)\citenamefont
  {Schmidt}, \citenamefont {Lafuente},\ and\ \citenamefont
  {Cuesta}}]{Schmidt/etal:2003}%
  \BibitemOpen
  \bibfield  {author} {\bibinfo {author} {\bibfnamefont {M.}~\bibnamefont
  {Schmidt}}, \bibinfo {author} {\bibfnamefont {L.}~\bibnamefont {Lafuente}}, \
  and\ \bibinfo {author} {\bibfnamefont {J.~A.}\ \bibnamefont {Cuesta}},\
  }\href {\doibase {10.1088/0953-8984/15/27/304}} {\bibfield  {journal}
  {\bibinfo  {journal} {J. Phys.: Condens. Matter}\ }\textbf {\bibinfo {volume}
  {15}},\ \bibinfo {pages} {4695} (\bibinfo {year} {2003})}\BibitemShut
  {NoStop}%
\bibitem [{\citenamefont {Lafuente}\ and\ \citenamefont
  {Cuesta}(2003{\natexlab{a}})}]{Lafuente/Cuesta:2003a}%
  \BibitemOpen
  \bibfield  {author} {\bibinfo {author} {\bibfnamefont {L.}~\bibnamefont
  {Lafuente}}\ and\ \bibinfo {author} {\bibfnamefont {J.~A.}\ \bibnamefont
  {Cuesta}},\ }\href {\doibase 10.1063/1.1615511} {\bibfield  {journal}
  {\bibinfo  {journal} {J. Chem. Phys.}\ }\textbf {\bibinfo {volume} {119}},\
  \bibinfo {pages} {10832} (\bibinfo {year} {2003}{\natexlab{a}})}\BibitemShut
  {NoStop}%
\bibitem [{\citenamefont {Lafuente}\ and\ \citenamefont
  {Cuesta}(2003{\natexlab{b}})}]{Lafuente/Cuesta:2003b}%
  \BibitemOpen
  \bibfield  {author} {\bibinfo {author} {\bibfnamefont {L.}~\bibnamefont
  {Lafuente}}\ and\ \bibinfo {author} {\bibfnamefont {J.~A.}\ \bibnamefont
  {Cuesta}},\ }\href {\doibase {10.1103/PhysRevE.68.066120}} {\bibfield
  {journal} {\bibinfo  {journal} {Phys. Rev. E}\ }\textbf {\bibinfo {volume}
  {68}},\ \bibinfo {pages} {066120} (\bibinfo {year}
  {2003}{\natexlab{b}})}\BibitemShut {NoStop}%
\bibitem [{\citenamefont {Lafuente}\ and\ \citenamefont
  {Cuesta}(2004)}]{Lafuente/Cuesta:2004}%
  \BibitemOpen
  \bibfield  {author} {\bibinfo {author} {\bibfnamefont {L.}~\bibnamefont
  {Lafuente}}\ and\ \bibinfo {author} {\bibfnamefont {J.~A.}\ \bibnamefont
  {Cuesta}},\ }\href {\doibase {10.1103/PhysRevLett.93.130603}} {\bibfield
  {journal} {\bibinfo  {journal} {Phys. Rev. Lett.}\ }\textbf {\bibinfo
  {volume} {93}},\ \bibinfo {pages} {130603} (\bibinfo {year}
  {2004})}\BibitemShut {NoStop}%
\bibitem [{\citenamefont {Heinrichs}\ \emph {et~al.}(2004)\citenamefont
  {Heinrichs}, \citenamefont {Dieterich}, \citenamefont {Frisch},\ and\
  \citenamefont {Maass}}]{Heinrichs/etal:2004}%
  \BibitemOpen
  \bibfield  {author} {\bibinfo {author} {\bibfnamefont {S.}~\bibnamefont
  {Heinrichs}}, \bibinfo {author} {\bibfnamefont {W.}~\bibnamefont
  {Dieterich}}, \bibinfo {author} {\bibfnamefont {H.~L.}\ \bibnamefont
  {Frisch}}, \ and\ \bibinfo {author} {\bibfnamefont {P.}~\bibnamefont
  {Maass}},\ }\href {\doibase 10.1023/B:JOSS.0000012518.27400.2a} {\bibfield
  {journal} {\bibinfo  {journal} {J. Stat. Phys.}\ }\textbf {\bibinfo {volume}
  {114}},\ \bibinfo {pages} {1115} (\bibinfo {year} {2004})}\BibitemShut
  {NoStop}%
\bibitem [{\citenamefont {Kessler}\ \emph {et~al.}(2002)\citenamefont
  {Kessler}, \citenamefont {Dieterich}, \citenamefont {Frisch}, \citenamefont
  {Gouyet},\ and\ \citenamefont {Maass}}]{Kessler/etal:2002}%
  \BibitemOpen
  \bibfield  {author} {\bibinfo {author} {\bibfnamefont {M.}~\bibnamefont
  {Kessler}}, \bibinfo {author} {\bibfnamefont {W.}~\bibnamefont {Dieterich}},
  \bibinfo {author} {\bibfnamefont {H.~L.}\ \bibnamefont {Frisch}}, \bibinfo
  {author} {\bibfnamefont {J.~F.}\ \bibnamefont {Gouyet}}, \ and\ \bibinfo
  {author} {\bibfnamefont {P.}~\bibnamefont {Maass}},\ }\href {\doibase
  10.1103/PhysRevE.65.066112} {\bibfield  {journal} {\bibinfo  {journal} {Phys.
  Rev. E}\ }\textbf {\bibinfo {volume} {65}},\ \bibinfo {pages} {066112}
  (\bibinfo {year} {2002})}\BibitemShut {NoStop}%
\bibitem [{\citenamefont {Dierl}\ \emph {et~al.}(2011)\citenamefont {Dierl},
  \citenamefont {Maass},\ and\ \citenamefont {Einax}}]{Dierl/etal:2011}%
  \BibitemOpen
  \bibfield  {author} {\bibinfo {author} {\bibfnamefont {M.}~\bibnamefont
  {Dierl}}, \bibinfo {author} {\bibfnamefont {P.}~\bibnamefont {Maass}}, \ and\
  \bibinfo {author} {\bibfnamefont {M.}~\bibnamefont {Einax}},\ }\href
  {\doibase {10.1209/0295-5075/93/50003}} {\bibfield  {journal} {\bibinfo
  {journal} {Europhys. Lett.}\ }\textbf {\bibinfo {volume} {93}},\ \bibinfo
  {pages} {50003} (\bibinfo {year} {2011})}\BibitemShut {NoStop}%
\bibitem [{\citenamefont {Dierl}\ \emph {et~al.}(2012)\citenamefont {Dierl},
  \citenamefont {Maass},\ and\ \citenamefont {Einax}}]{Dierl/etal:2012}%
  \BibitemOpen
  \bibfield  {author} {\bibinfo {author} {\bibfnamefont {M.}~\bibnamefont
  {Dierl}}, \bibinfo {author} {\bibfnamefont {P.}~\bibnamefont {Maass}}, \ and\
  \bibinfo {author} {\bibfnamefont {M.}~\bibnamefont {Einax}},\ }\href
  {\doibase 10.1103/PhysRevLett.108.060603} {\bibfield  {journal} {\bibinfo
  {journal} {Phys. Rev. Lett.}\ }\textbf {\bibinfo {volume} {108}},\ \bibinfo
  {pages} {060603} (\bibinfo {year} {2012})}\BibitemShut {NoStop}%
\bibitem [{\citenamefont {Baxter}(1980)}]{Baxter:1980}%
  \BibitemOpen
  \bibfield  {author} {\bibinfo {author} {\bibfnamefont {R.~J.}\ \bibnamefont
  {Baxter}},\ }\href {http://stacks.iop.org/0305-4470/13/i=3/a=007} {\bibfield
  {journal} {\bibinfo  {journal} {J. Phys. A: Math. Gen.}\ }\textbf {\bibinfo
  {volume} {13}},\ \bibinfo {pages} {L61} (\bibinfo {year} {1980})}\BibitemShut
  {NoStop}%
\bibitem [{\citenamefont {Percus}(1982)}]{Percus:1982}%
  \BibitemOpen
  \bibfield  {author} {\bibinfo {author} {\bibfnamefont {J.~K.}\ \bibnamefont
  {Percus}},\ }\href {\doibase 10.1007/BF01011623} {\bibfield  {journal}
  {\bibinfo  {journal} {J. Stat. Phys.}\ }\textbf {\bibinfo {volume} {28}},\
  \bibinfo {pages} {67} (\bibinfo {year} {1982})}\BibitemShut {NoStop}%
\bibitem [{\citenamefont {Brannock}\ and\ \citenamefont
  {Percus}(1996)}]{Brannock/Percus:1996}%
  \BibitemOpen
  \bibfield  {author} {\bibinfo {author} {\bibfnamefont {G.~R.}\ \bibnamefont
  {Brannock}}\ and\ \bibinfo {author} {\bibfnamefont {J.~K.}\ \bibnamefont
  {Percus}},\ }\href {\doibase 10.1063/1.471920} {\bibfield  {journal}
  {\bibinfo  {journal} {J. Chem. Phys.}\ }\textbf {\bibinfo {volume} {105}},\
  \bibinfo {pages} {614} (\bibinfo {year} {1996})}\BibitemShut {NoStop}%
\bibitem [{\citenamefont {Choudhury}\ and\ \citenamefont
  {Ghosh}(1997)}]{Choudhury/Ghosh:1997}%
  \BibitemOpen
  \bibfield  {author} {\bibinfo {author} {\bibfnamefont {N.}~\bibnamefont
  {Choudhury}}\ and\ \bibinfo {author} {\bibfnamefont {S.~K.}\ \bibnamefont
  {Ghosh}},\ }\href {\doibase 10.1063/1.473286} {\bibfield  {journal} {\bibinfo
   {journal} {J. Chem. Phys.}\ }\textbf {\bibinfo {volume} {106}},\ \bibinfo
  {pages} {1576} (\bibinfo {year} {1997})}\BibitemShut {NoStop}%
\bibitem [{\citenamefont {Acedo}\ and\ \citenamefont
  {Santos}(2001)}]{Acedo/Santos:2001}%
  \BibitemOpen
  \bibfield  {author} {\bibinfo {author} {\bibfnamefont {L.}~\bibnamefont
  {Acedo}}\ and\ \bibinfo {author} {\bibfnamefont {A.}~\bibnamefont {Santos}},\
  }\href {\doibase 10.1063/1.1384419} {\bibfield  {journal} {\bibinfo
  {journal} {J. Chem. Phys.}\ }\textbf {\bibinfo {volume} {115}},\ \bibinfo
  {pages} {2805} (\bibinfo {year} {2001})}\BibitemShut {NoStop}%
\bibitem [{\citenamefont {Gazzillo}\ and\ \citenamefont
  {Giacometti}(2004)}]{Gazzillo/etal:2004}%
  \BibitemOpen
  \bibfield  {author} {\bibinfo {author} {\bibfnamefont {D.}~\bibnamefont
  {Gazzillo}}\ and\ \bibinfo {author} {\bibfnamefont {A.}~\bibnamefont
  {Giacometti}},\ }\href {\doibase 10.1063/1.1645781} {\bibfield  {journal}
  {\bibinfo  {journal} {J. Chem. Phys.}\ }\textbf {\bibinfo {volume} {120}},\
  \bibinfo {pages} {4742} (\bibinfo {year} {2004})}\BibitemShut {NoStop}%
\bibitem [{\citenamefont {Miller}\ and\ \citenamefont
  {Frenkel}(2004{\natexlab{a}})}]{Miller/Frenkel:2004a}%
  \BibitemOpen
  \bibfield  {author} {\bibinfo {author} {\bibfnamefont {M.~A.}\ \bibnamefont
  {Miller}}\ and\ \bibinfo {author} {\bibfnamefont {D.}~\bibnamefont
  {Frenkel}},\ }\href {\doibase 10.1063/1.1758693} {\bibfield  {journal}
  {\bibinfo  {journal} {J. Chem. Phys.}\ }\textbf {\bibinfo {volume} {121}},\
  \bibinfo {pages} {535} (\bibinfo {year} {2004}{\natexlab{a}})}\BibitemShut
  {NoStop}%
\bibitem [{\citenamefont {Miller}\ and\ \citenamefont
  {Frenkel}(2004{\natexlab{b}})}]{Miller/Frenkel:2004b}%
  \BibitemOpen
  \bibfield  {author} {\bibinfo {author} {\bibfnamefont {M.~A.}\ \bibnamefont
  {Miller}}\ and\ \bibinfo {author} {\bibfnamefont {D.}~\bibnamefont
  {Frenkel}},\ }\href@noop {} {\bibfield  {journal} {\bibinfo  {journal} {J.
  Phys.:Condens. Matter}\ }\textbf {\bibinfo {volume} {16}},\ \bibinfo {pages}
  {S4901} (\bibinfo {year} {2004}{\natexlab{b}})}\BibitemShut {NoStop}%
\bibitem [{\citenamefont {Rickayzen}\ and\ \citenamefont
  {Heyes}(2007)}]{Rickayzen/Heyes:2007}%
  \BibitemOpen
  \bibfield  {author} {\bibinfo {author} {\bibfnamefont {G.}~\bibnamefont
  {Rickayzen}}\ and\ \bibinfo {author} {\bibfnamefont {D.~M.}\ \bibnamefont
  {Heyes}},\ }\href {\doibase 10.1063/1.2647150} {\bibfield  {journal}
  {\bibinfo  {journal} {J. Chem. Phys.}\ }\textbf {\bibinfo {volume} {126}},\
  \bibinfo {eid} {114504} (\bibinfo {year} {2007})}\BibitemShut {NoStop}%
\bibitem [{\citenamefont {Buzzaccaro}\ \emph {et~al.}(2007)\citenamefont
  {Buzzaccaro}, \citenamefont {Rusconi},\ and\ \citenamefont
  {Piazza}}]{Buzzaccaro/etal:2007}%
  \BibitemOpen
  \bibfield  {author} {\bibinfo {author} {\bibfnamefont {S.}~\bibnamefont
  {Buzzaccaro}}, \bibinfo {author} {\bibfnamefont {R.}~\bibnamefont {Rusconi}},
  \ and\ \bibinfo {author} {\bibfnamefont {R.}~\bibnamefont {Piazza}},\ }\href
  {\doibase 10.1103/PhysRevLett.99.098301} {\bibfield  {journal} {\bibinfo
  {journal} {Phys. Rev. Lett.}\ }\textbf {\bibinfo {volume} {99}},\ \bibinfo
  {pages} {098301} (\bibinfo {year} {2007})}\BibitemShut {NoStop}%
\bibitem [{\citenamefont {Santos}\ \emph {et~al.}(2008)\citenamefont {Santos},
  \citenamefont {Fantoni},\ and\ \citenamefont
  {Giacometti}}]{Santos/etal:2008}%
  \BibitemOpen
  \bibfield  {author} {\bibinfo {author} {\bibfnamefont {A.}~\bibnamefont
  {Santos}}, \bibinfo {author} {\bibfnamefont {R.}~\bibnamefont {Fantoni}}, \
  and\ \bibinfo {author} {\bibfnamefont {A.}~\bibnamefont {Giacometti}},\
  }\href {\doibase 10.1103/PhysRevE.77.051206} {\bibfield  {journal} {\bibinfo
  {journal} {Phys. Rev. E}\ }\textbf {\bibinfo {volume} {77}},\ \bibinfo
  {pages} {051206} (\bibinfo {year} {2008})}\BibitemShut {NoStop}%
\bibitem [{\citenamefont {Hansen-Goos}\ and\ \citenamefont
  {Wettlaufer}(2011)}]{Hansen-goos/Wettlaufer:2011}%
  \BibitemOpen
  \bibfield  {author} {\bibinfo {author} {\bibfnamefont {H.}~\bibnamefont
  {Hansen-Goos}}\ and\ \bibinfo {author} {\bibfnamefont {J.~S.}\ \bibnamefont
  {Wettlaufer}},\ }\href {\doibase 10.1063/1.3528226} {\bibfield  {journal}
  {\bibinfo  {journal} {J. Chem. Phys.}\ }\textbf {\bibinfo {volume} {134}},\
  \bibinfo {eid} {014506} (\bibinfo {year} {2011})}\BibitemShut {NoStop}%
\bibitem [{\citenamefont {Hansen-Goos}\ \emph {et~al.}(2012)\citenamefont
  {Hansen-Goos}, \citenamefont {Miller},\ and\ \citenamefont
  {Wettlaufer}}]{Hansen-goos/etal:2012}%
  \BibitemOpen
  \bibfield  {author} {\bibinfo {author} {\bibfnamefont {H.}~\bibnamefont
  {Hansen-Goos}}, \bibinfo {author} {\bibfnamefont {M.~A.}\ \bibnamefont
  {Miller}}, \ and\ \bibinfo {author} {\bibfnamefont {J.~S.}\ \bibnamefont
  {Wettlaufer}},\ }\href {\doibase 10.1103/PhysRevLett.108.047801} {\bibfield
  {journal} {\bibinfo  {journal} {Phys. Rev. Lett.}\ }\textbf {\bibinfo
  {volume} {108}},\ \bibinfo {pages} {047801} (\bibinfo {year}
  {2012})}\BibitemShut {NoStop}%
\bibitem [{\citenamefont {Jamnik}(1998)}]{Jamnik:1998}%
  \BibitemOpen
  \bibfield  {author} {\bibinfo {author} {\bibfnamefont {A.}~\bibnamefont
  {Jamnik}},\ }\href {\doibase 10.1063/1.477746} {\bibfield  {journal}
  {\bibinfo  {journal} {J. Chem. Phys.}\ }\textbf {\bibinfo {volume} {109}},\
  \bibinfo {pages} {11085} (\bibinfo {year} {1998})}\BibitemShut {NoStop}%
\bibitem [{\citenamefont {Pontoni}\ \emph {et~al.}(2003)\citenamefont
  {Pontoni}, \citenamefont {Finet}, \citenamefont {Narayanan},\ and\
  \citenamefont {Rennie}}]{Pontoni/etal:2003}%
  \BibitemOpen
  \bibfield  {author} {\bibinfo {author} {\bibfnamefont {D.}~\bibnamefont
  {Pontoni}}, \bibinfo {author} {\bibfnamefont {S.}~\bibnamefont {Finet}},
  \bibinfo {author} {\bibfnamefont {T.}~\bibnamefont {Narayanan}}, \ and\
  \bibinfo {author} {\bibfnamefont {A.~R.}\ \bibnamefont {Rennie}},\ }\href
  {\doibase 10.1063/1.1601605} {\bibfield  {journal} {\bibinfo  {journal} {J.
  Chem. Phys.}\ }\textbf {\bibinfo {volume} {119}},\ \bibinfo {pages} {6157}
  (\bibinfo {year} {2003})}\BibitemShut {NoStop}%
\bibitem [{\citenamefont {Lajovic}\ \emph {et~al.}(2009)\citenamefont
  {Lajovic}, \citenamefont {Tomsic},\ and\ \citenamefont
  {Jamnik}}]{Lajovic/etal:2009}%
  \BibitemOpen
  \bibfield  {author} {\bibinfo {author} {\bibfnamefont {A.}~\bibnamefont
  {Lajovic}}, \bibinfo {author} {\bibfnamefont {M.}~\bibnamefont {Tomsic}}, \
  and\ \bibinfo {author} {\bibfnamefont {A.}~\bibnamefont {Jamnik}},\ }\href
  {\doibase 10.1063/1.3081144} {\bibfield  {journal} {\bibinfo  {journal} {J.
  Chem. Phys.}\ }\textbf {\bibinfo {volume} {130}},\ \bibinfo {eid} {104101}
  (\bibinfo {year} {2009})}\BibitemShut {NoStop}%
\bibitem [{\citenamefont {Dasmahapatra}\ \emph {et~al.}(2009)\citenamefont
  {Dasmahapatra}, \citenamefont {Nanavati},\ and\ \citenamefont
  {Kumaraswamy}}]{Dasmahapatra/etal:2009}%
  \BibitemOpen
  \bibfield  {author} {\bibinfo {author} {\bibfnamefont {A.~K.}\ \bibnamefont
  {Dasmahapatra}}, \bibinfo {author} {\bibfnamefont {H.}~\bibnamefont
  {Nanavati}}, \ and\ \bibinfo {author} {\bibfnamefont {G.}~\bibnamefont
  {Kumaraswamy}},\ }\href {\doibase 10.1063/1.3174449} {\bibfield  {journal}
  {\bibinfo  {journal} {J. Chem. Phys.}\ }\textbf {\bibinfo {volume} {131}},\
  \bibinfo {eid} {074905} (\bibinfo {year} {2009})}\BibitemShut {NoStop}%
\bibitem [{\citenamefont {Hoy}\ and\ \citenamefont
  {O'Hern}(2010)}]{Hoy/OHern:2010}%
  \BibitemOpen
  \bibfield  {author} {\bibinfo {author} {\bibfnamefont {R.~S.}\ \bibnamefont
  {Hoy}}\ and\ \bibinfo {author} {\bibfnamefont {C.~S.}\ \bibnamefont
  {O'Hern}},\ }\href {\doibase 10.1103/PhysRevLett.105.068001} {\bibfield
  {journal} {\bibinfo  {journal} {Phys. Rev. Lett.}\ }\textbf {\bibinfo
  {volume} {105}},\ \bibinfo {pages} {068001} (\bibinfo {year}
  {2010})}\BibitemShut {NoStop}%
\bibitem [{\citenamefont {Amokrane}\ and\ \citenamefont
  {Regnaut}(1997)}]{Amokrane/Regnaut:1997}%
  \BibitemOpen
  \bibfield  {author} {\bibinfo {author} {\bibfnamefont {S.}~\bibnamefont
  {Amokrane}}\ and\ \bibinfo {author} {\bibfnamefont {C.}~\bibnamefont
  {Regnaut}},\ }\href {\doibase 10.1063/1.473201} {\bibfield  {journal}
  {\bibinfo  {journal} {J. Chem. Phys.}\ }\textbf {\bibinfo {volume} {106}},\
  \bibinfo {pages} {376} (\bibinfo {year} {1997})}\BibitemShut {NoStop}%
\bibitem [{\citenamefont {Lomakin}\ \emph {et~al.}(1996)\citenamefont
  {Lomakin}, \citenamefont {Asherie},\ and\ \citenamefont
  {Benedek}}]{Lomakin/etal:1996}%
  \BibitemOpen
  \bibfield  {author} {\bibinfo {author} {\bibfnamefont {A.}~\bibnamefont
  {Lomakin}}, \bibinfo {author} {\bibfnamefont {N.}~\bibnamefont {Asherie}}, \
  and\ \bibinfo {author} {\bibfnamefont {G.~B.}\ \bibnamefont {Benedek}},\
  }\href {\doibase 10.1063/1.470751} {\bibfield  {journal} {\bibinfo  {journal}
  {J. Chem. Phys.}\ }\textbf {\bibinfo {volume} {104}},\ \bibinfo {pages}
  {1646} (\bibinfo {year} {1996})}\BibitemShut {NoStop}%
\bibitem [{\citenamefont {Braun}(2002)}]{Braun:2002}%
  \BibitemOpen
  \bibfield  {author} {\bibinfo {author} {\bibfnamefont {F.~N.}\ \bibnamefont
  {Braun}},\ }\href {\doibase 10.1063/1.1461358} {\bibfield  {journal}
  {\bibinfo  {journal} {J. Chem. Phys.}\ }\textbf {\bibinfo {volume} {116}},\
  \bibinfo {pages} {6826} (\bibinfo {year} {2002})}\BibitemShut {NoStop}%
\bibitem [{\citenamefont {Xu}\ \emph {et~al.}(2011)\citenamefont {Xu},
  \citenamefont {Feng}, \citenamefont {Sha}, \citenamefont {Seeman},\ and\
  \citenamefont {Chaikin}}]{Xu/etal:2011}%
  \BibitemOpen
  \bibfield  {author} {\bibinfo {author} {\bibfnamefont {Q.}~\bibnamefont
  {Xu}}, \bibinfo {author} {\bibfnamefont {L.}~\bibnamefont {Feng}}, \bibinfo
  {author} {\bibfnamefont {R.}~\bibnamefont {Sha}}, \bibinfo {author}
  {\bibfnamefont {N.~C.}\ \bibnamefont {Seeman}}, \ and\ \bibinfo {author}
  {\bibfnamefont {P.~M.}\ \bibnamefont {Chaikin}},\ }\href {\doibase
  10.1103/PhysRevLett.106.228102} {\bibfield  {journal} {\bibinfo  {journal}
  {Phys. Rev. Lett.}\ }\textbf {\bibinfo {volume} {106}},\ \bibinfo {pages}
  {228102} (\bibinfo {year} {2011})}\BibitemShut {NoStop}%
\bibitem [{\citenamefont {Dreyfus}\ \emph {et~al.}(2009)\citenamefont
  {Dreyfus}, \citenamefont {Leunissen}, \citenamefont {Sha}, \citenamefont
  {Tkachenko}, \citenamefont {Seeman}, \citenamefont {Pine},\ and\
  \citenamefont {Chaikin}}]{Dreyfus/etal:2009}%
  \BibitemOpen
  \bibfield  {author} {\bibinfo {author} {\bibfnamefont {R.}~\bibnamefont
  {Dreyfus}}, \bibinfo {author} {\bibfnamefont {M.~E.}\ \bibnamefont
  {Leunissen}}, \bibinfo {author} {\bibfnamefont {R.}~\bibnamefont {Sha}},
  \bibinfo {author} {\bibfnamefont {A.~V.}\ \bibnamefont {Tkachenko}}, \bibinfo
  {author} {\bibfnamefont {N.~C.}\ \bibnamefont {Seeman}}, \bibinfo {author}
  {\bibfnamefont {D.~J.}\ \bibnamefont {Pine}}, \ and\ \bibinfo {author}
  {\bibfnamefont {P.~M.}\ \bibnamefont {Chaikin}},\ }\href {\doibase
  10.1103/PhysRevLett.102.048301} {\bibfield  {journal} {\bibinfo  {journal}
  {Phys. Rev. Lett.}\ }\textbf {\bibinfo {volume} {102}},\ \bibinfo {pages}
  {048301} (\bibinfo {year} {2009})}\BibitemShut {NoStop}%
\bibitem [{\citenamefont {Stell}(1995)}]{Stell:1995}%
  \BibitemOpen
  \bibfield  {author} {\bibinfo {author} {\bibfnamefont {G.}~\bibnamefont
  {Stell}},\ }\href {\doibase 10.1007/BF02183346} {\bibfield  {journal}
  {\bibinfo  {journal} {J. Stat. Phys}\ }\textbf {\bibinfo {volume} {78}},\
  \bibinfo {pages} {197} (\bibinfo {year} {1995})}\BibitemShut {NoStop}%
\bibitem [{\citenamefont {Seto}\ \emph {et~al.}(2000)\citenamefont {Seto},
  \citenamefont {Okuhara}, \citenamefont {Kawabata}, \citenamefont {Takeda},
  \citenamefont {Nagao}, \citenamefont {Suzuki}, \citenamefont {Kamikubo},\
  and\ \citenamefont {Amemiya}}]{Seto/etal:2000}%
  \BibitemOpen
  \bibfield  {author} {\bibinfo {author} {\bibfnamefont {H.}~\bibnamefont
  {Seto}}, \bibinfo {author} {\bibfnamefont {D.}~\bibnamefont {Okuhara}},
  \bibinfo {author} {\bibfnamefont {Y.}~\bibnamefont {Kawabata}}, \bibinfo
  {author} {\bibfnamefont {T.}~\bibnamefont {Takeda}}, \bibinfo {author}
  {\bibfnamefont {M.}~\bibnamefont {Nagao}}, \bibinfo {author} {\bibfnamefont
  {J.}~\bibnamefont {Suzuki}}, \bibinfo {author} {\bibfnamefont
  {H.}~\bibnamefont {Kamikubo}}, \ and\ \bibinfo {author} {\bibfnamefont
  {Y.}~\bibnamefont {Amemiya}},\ }\href {\doibase 10.1063/1.481695} {\bibfield
  {journal} {\bibinfo  {journal} {J. Chem. Phys.}\ }\textbf {\bibinfo {volume}
  {112}},\ \bibinfo {pages} {10608} (\bibinfo {year} {2000})}\BibitemShut
  {NoStop}%
\bibitem [{\citenamefont {Robertus}\ \emph {et~al.}(1989)\citenamefont
  {Robertus}, \citenamefont {Philipse}, \citenamefont {Joosten},\ and\
  \citenamefont {Levine}}]{Robertus/etal:1989}%
  \BibitemOpen
  \bibfield  {author} {\bibinfo {author} {\bibfnamefont {C.}~\bibnamefont
  {Robertus}}, \bibinfo {author} {\bibfnamefont {W.~H.}\ \bibnamefont
  {Philipse}}, \bibinfo {author} {\bibfnamefont {J.~G.~H.}\ \bibnamefont
  {Joosten}}, \ and\ \bibinfo {author} {\bibfnamefont {Y.~K.}\ \bibnamefont
  {Levine}},\ }\href {\doibase 10.1063/1.456635} {\bibfield  {journal}
  {\bibinfo  {journal} {J. Chem. Phys.}\ }\textbf {\bibinfo {volume} {90}},\
  \bibinfo {pages} {4482} (\bibinfo {year} {1989})}\BibitemShut {NoStop}%
\bibitem [{\citenamefont {Gazzillo}\ \emph {et~al.}(2006)\citenamefont
  {Gazzillo}, \citenamefont {Giacometti}, \citenamefont {Fantoni},\ and\
  \citenamefont {Sollich}}]{Gazzillo/etal:2006}%
  \BibitemOpen
  \bibfield  {author} {\bibinfo {author} {\bibfnamefont {D.}~\bibnamefont
  {Gazzillo}}, \bibinfo {author} {\bibfnamefont {A.}~\bibnamefont
  {Giacometti}}, \bibinfo {author} {\bibfnamefont {R.}~\bibnamefont {Fantoni}},
  \ and\ \bibinfo {author} {\bibfnamefont {P.}~\bibnamefont {Sollich}},\ }\href
  {\doibase 10.1103/PhysRevE.74.051407} {\bibfield  {journal} {\bibinfo
  {journal} {Phys. Rev. E}\ }\textbf {\bibinfo {volume} {74}},\ \bibinfo
  {pages} {051407} (\bibinfo {year} {2006})}\BibitemShut {NoStop}%
\bibitem [{\citenamefont {Buschle}\ \emph
  {et~al.}(2000{\natexlab{a}})\citenamefont {Buschle}, \citenamefont {Maass},\
  and\ \citenamefont {Dieterich}}]{Buschle/etal:2000a}%
  \BibitemOpen
  \bibfield  {author} {\bibinfo {author} {\bibfnamefont {J.}~\bibnamefont
  {Buschle}}, \bibinfo {author} {\bibfnamefont {P.}~\bibnamefont {Maass}}, \
  and\ \bibinfo {author} {\bibfnamefont {W.}~\bibnamefont {Dieterich}},\
  }\href@noop {} {\bibfield  {journal} {\bibinfo  {journal} {J. Phys. A}\
  }\textbf {\bibinfo {volume} {33}},\ \bibinfo {pages} {L41} (\bibinfo {year}
  {2000}{\natexlab{a}})}\BibitemShut {NoStop}%
\bibitem [{\citenamefont {Buschle}\ \emph
  {et~al.}(2000{\natexlab{b}})\citenamefont {Buschle}, \citenamefont {Maass},\
  and\ \citenamefont {Dieterich}}]{Buschle/etal:2000b}%
  \BibitemOpen
  \bibfield  {author} {\bibinfo {author} {\bibfnamefont {J.}~\bibnamefont
  {Buschle}}, \bibinfo {author} {\bibfnamefont {P.}~\bibnamefont {Maass}}, \
  and\ \bibinfo {author} {\bibfnamefont {W.}~\bibnamefont {Dieterich}},\ }\href
  {\doibase 10.1023/A:1018652808652} {\bibfield  {journal} {\bibinfo  {journal}
  {J. Stat. Phys.}\ }\textbf {\bibinfo {volume} {99}},\ \bibinfo {pages} {273}
  (\bibinfo {year} {2000}{\natexlab{b}})}\BibitemShut {NoStop}%
\bibitem [{\citenamefont {Baxter}(1968)}]{Baxter:1968}%
  \BibitemOpen
  \bibfield  {author} {\bibinfo {author} {\bibfnamefont {R.~J.}\ \bibnamefont
  {Baxter}},\ }\href {\doibase 10.1063/1.1670482} {\bibfield  {journal}
  {\bibinfo  {journal} {J. Chem. Phys.}\ }\textbf {\bibinfo {volume} {49}},\
  \bibinfo {pages} {2770} (\bibinfo {year} {1968})}\BibitemShut {NoStop}%
\bibitem [{\citenamefont {Abe}(1959)}]{Abe:1959}%
  \BibitemOpen
  \bibfield  {author} {\bibinfo {author} {\bibfnamefont {R.}~\bibnamefont
  {Abe}},\ }\href {\doibase 10.1143/PTP.21.421} {\bibfield  {journal} {\bibinfo
   {journal} {Prog. Theor. Phys.}\ }\textbf {\bibinfo {volume} {21}},\ \bibinfo
  {pages} {421} (\bibinfo {year} {1959})}\BibitemShut {NoStop}%
\bibitem [{\citenamefont {Salsburg}\ \emph {et~al.}(1953)\citenamefont
  {Salsburg}, \citenamefont {Zwanzig},\ and\ \citenamefont
  {Kirkwood}}]{Salsburg/etal:1953}%
  \BibitemOpen
  \bibfield  {author} {\bibinfo {author} {\bibfnamefont {Z.~W.}\ \bibnamefont
  {Salsburg}}, \bibinfo {author} {\bibfnamefont {R.~W.}\ \bibnamefont
  {Zwanzig}}, \ and\ \bibinfo {author} {\bibfnamefont {J.~G.}\ \bibnamefont
  {Kirkwood}},\ }\href {\doibase 10.1063/1.1699116} {\bibfield  {journal}
  {\bibinfo  {journal} {J. Chem. Phys.}\ }\textbf {\bibinfo {volume} {21}},\
  \bibinfo {pages} {1098} (\bibinfo {year} {1953})}\BibitemShut {NoStop}%
\bibitem [{\citenamefont {Mermin}(1965)}]{Mermin:1965}%
  \BibitemOpen
  \bibfield  {author} {\bibinfo {author} {\bibfnamefont {N.~D.}\ \bibnamefont
  {Mermin}},\ }\href {\doibase 10.1103/PhysRev.137.A1441} {\bibfield  {journal}
  {\bibinfo  {journal} {Phys. Rev.}\ }\textbf {\bibinfo {volume} {137}},\
  \bibinfo {pages} {A1441} (\bibinfo {year} {1965})}\BibitemShut {NoStop}%
\bibitem [{\citenamefont {Lafuente}\ and\ \citenamefont
  {Cuesta}(2005)}]{Lafuente/Cuesta:2005}%
  \BibitemOpen
  \bibfield  {author} {\bibinfo {author} {\bibfnamefont {L.}~\bibnamefont
  {Lafuente}}\ and\ \bibinfo {author} {\bibfnamefont {J.~A.}\ \bibnamefont
  {Cuesta}},\ }\href@noop {} {\bibfield  {journal} {\bibinfo  {journal} {J.
  Phys A: Math. Gen.}\ }\textbf {\bibinfo {volume} {38}},\ \bibinfo {pages}
  {7461} (\bibinfo {year} {2005})}\BibitemShut {NoStop}%
\bibitem [{\citenamefont {Gundlach}\ \emph {et~al.}(2013)\citenamefont
  {Gundlach}, \citenamefont {Karbach}, \citenamefont {Liu},\ and\ \citenamefont
  {M{\"u}ller}}]{Gundlach/etal:2013}%
  \BibitemOpen
  \bibfield  {author} {\bibinfo {author} {\bibfnamefont {N.}~\bibnamefont
  {Gundlach}}, \bibinfo {author} {\bibfnamefont {M.}~\bibnamefont {Karbach}},
  \bibinfo {author} {\bibfnamefont {D.}~\bibnamefont {Liu}}, \ and\ \bibinfo
  {author} {\bibfnamefont {G.}~\bibnamefont {M{\"u}ller}},\ }\href@noop {}
  {\bibfield  {journal} {\bibinfo  {journal} {J. Stat. Mech. P04018}\ }
  (\bibinfo {year} {2013})}\BibitemShut {NoStop}%
\bibitem [{\citenamefont {Percus}(1989)}]{Percus:1989}%
  \BibitemOpen
  \bibfield  {author} {\bibinfo {author} {\bibfnamefont {J.~K.}\ \bibnamefont
  {Percus}},\ }\href {http://stacks.iop.org/0953-8984/1/i=17/a=011} {\bibfield
  {journal} {\bibinfo  {journal} {J. Phys: Condens. Matter}\ }\textbf {\bibinfo
  {volume} {1}},\ \bibinfo {pages} {2911} (\bibinfo {year} {1989})}\BibitemShut
  {NoStop}%
\bibitem [{\citenamefont {Berlin}\ and\ \citenamefont
  {Kac.}(1952)}]{Berlin/Kac:1952}%
  \BibitemOpen
  \bibfield  {author} {\bibinfo {author} {\bibfnamefont {T.~H.}\ \bibnamefont
  {Berlin}}\ and\ \bibinfo {author} {\bibfnamefont {M.}~\bibnamefont {Kac.}},\
  }\href {\doibase 10.1103/PhysRev.86.821} {\bibfield  {journal} {\bibinfo
  {journal} {Phys. Rev.}\ }\textbf {\bibinfo {volume} {86}},\ \bibinfo {pages}
  {821} (\bibinfo {year} {1952})}\BibitemShut {NoStop}%
\bibitem [{\citenamefont {Mart{\'i}nez-Rat{\'o}n}(2007)}]{Martinez:2007}%
  \BibitemOpen
  \bibfield  {author} {\bibinfo {author} {\bibfnamefont {Y.}~\bibnamefont
  {Mart{\'i}nez-Rat{\'o}n}},\ }\href {\doibase 10.1103/PhysRevE.75.051708}
  {\bibfield  {journal} {\bibinfo  {journal} {Phys. Rev. E}\ }\textbf {\bibinfo
  {volume} {75}},\ \bibinfo {pages} {051708} (\bibinfo {year}
  {2007})}\BibitemShut {NoStop}%
\bibitem [{\citenamefont {Mart{\'i}nez-Rat{\'o}n}\ \emph
  {et~al.}(2008)\citenamefont {Mart{\'i}nez-Rat{\'o}n}, \citenamefont {Varga},\
  and\ \citenamefont {Velasco}}]{Martinez/etal:2008}%
  \BibitemOpen
  \bibfield  {author} {\bibinfo {author} {\bibfnamefont {Y.}~\bibnamefont
  {Mart{\'i}nez-Rat{\'o}n}}, \bibinfo {author} {\bibfnamefont {S.}~\bibnamefont
  {Varga}}, \ and\ \bibinfo {author} {\bibfnamefont {E.}~\bibnamefont
  {Velasco}},\ }\href {\doibase 10.1103/PhysRevE.78.031705} {\bibfield
  {journal} {\bibinfo  {journal} {Phys. Rev. E}\ }\textbf {\bibinfo {volume}
  {78}},\ \bibinfo {pages} {031705} (\bibinfo {year} {2008})}\BibitemShut
  {NoStop}%
\bibitem [{\citenamefont {de~las Heras}\ \emph {et~al.}(2010)\citenamefont
  {de~las Heras}, \citenamefont {Mart{\'i}nez-Rat{\'o}n},\ and\ \citenamefont
  {Velasco}}]{Heras/etal:2010}%
  \BibitemOpen
  \bibfield  {author} {\bibinfo {author} {\bibfnamefont {D.}~\bibnamefont
  {de~las Heras}}, \bibinfo {author} {\bibfnamefont {Y.}~\bibnamefont
  {Mart{\'i}nez-Rat{\'o}n}}, \ and\ \bibinfo {author} {\bibfnamefont
  {E.}~\bibnamefont {Velasco}},\ }\href {\doibase 10.1103/PhysRevE.81.021706}
  {\bibfield  {journal} {\bibinfo  {journal} {Phys. Rev. E}\ }\textbf {\bibinfo
  {volume} {81}},\ \bibinfo {pages} {021706} (\bibinfo {year}
  {2010})}\BibitemShut {NoStop}%
\bibitem [{\citenamefont {Fisher}\ and\ \citenamefont
  {Widom}(1969)}]{Fisher/Widom:1969}%
  \BibitemOpen
  \bibfield  {author} {\bibinfo {author} {\bibfnamefont {M.~E.}\ \bibnamefont
  {Fisher}}\ and\ \bibinfo {author} {\bibfnamefont {B.}~\bibnamefont {Widom}},\
  }\href {\doibase 10.1063/1.1671624} {\bibfield  {journal} {\bibinfo
  {journal} {J. Chem. Phys.}\ }\textbf {\bibinfo {volume} {50}},\ \bibinfo
  {pages} {3756} (\bibinfo {year} {1969})}\BibitemShut {NoStop}%
\end{thebibliography}

%

\end{document}